\documentclass[11pt]{article}
\usepackage{graphicx}
\usepackage{amssymb}
\usepackage{amsmath} 
\usepackage{a4wide}
\usepackage[latin1]{inputenc}
\usepackage[T1]{fontenc}
\usepackage{soul}
\usepackage{color}
\usepackage{mdframed}
\usepackage{enumitem}
\usepackage{MnSymbol}
\usepackage{tikz}
\usetikzlibrary{arrows}
\usetikzlibrary{arrows.meta}
\usetikzlibrary{shapes,backgrounds}
\usepackage{tikz-3dplot,tikz-cd}
\tikzset{>=latex}

    \usetikzlibrary{positioning}
\def\d{\mathop{\mathrm{d}}}
\def\nat{\mathbb{{N}}}
\def\cA{\mathcal{{A}}}

\setul{0.4ex}{1ex}
\definecolor{thegrey}{rgb}{0.9, 0.9, 0.9}
\definecolor{Yellow}{rgb}{1.0, 1.0, 0.0}
\definecolor{brightgreen}{rgb}{0.45, 0.95, 0.0}
\newcommand{\gr}[1]{ {\sethlcolor{brightgreen} \hl{#1}} } 
\tikzstyle{default_conn} = [draw, -{Latex[scale=1.0]}, draw=black, line width=0.5]

\makeatletter
\DeclareRobustCommand{\bigmin}{%
  \mathop{\vphantom{\sum}\mathpalette\bigmin@\relax}\slimits@
}

\newcommand{\bigmin@}[2]{%
  \vcenter{%
    \sbox\z@{$#1\sum$}%
    \hbox{\resizebox{1.9\dimexpr\ht\z@+\dp\z@}{!}{$\m@th\min$}}%
  }%
}
\makeatother

\begin{document}

\author{J. Draeger}
\title{Some Remarks about the Complexity of Epidemics Management}
\date{}

\maketitle

\newcommand*\elevation{25}
\newcommand*\anglerot{300}
\pgfmathsetmacro\xc{cos(\anglerot)}  
\pgfmathsetmacro\xs{sin(\anglerot)}   
\pgfmathsetmacro\yc{cos(\elevation)} 
\pgfmathsetmacro\ys{sin(\elevation)} 
\newcommand*\axexthreed{\xs*1cm,-\xc*1cm} 
\newcommand*\axeythreed{\yc*1cm,-\ys*1cm}
\newcommand*\axezthreed{0cm,1cm} 
\tikzset{xy slant style/.style={blue, yslant=-tan(\elevation), xslant=-tan(\elevation+\anglerot),  yscale=cos(\anglerot), xscale=cos(\elevation), rotate=\anglerot}}

\begin{abstract}
Recent outbreaks of Ebola, H1N1 and other infectious diseases have 
shown that the assumptions underlying the established theory of
epidemics management are too idealistic. For an improvement of 
procedures and organizations involved in fighting epidemics,
extended models of epidemics management are required. The necessary
extensions consist in a representation of the management loop and 
the potential frictions influencing the loop. The effects of the
nondeterministic frictions can be taken into account by including the 
measures of robustness and risk in the assessment of management options.
Thus, besides of the increased structural complexity resulting from 
the model extensions, the computational complexity of the task of 
epidemics management --- interpreted as an optimization problem --- is 
increased as well. This is a serious obstacle for analyzing the model 
and may require an additional preprocessing enabling a simplification
of the analysis process. 
The paper closes with an outlook discussing some forthcoming problems.
\end{abstract}




\section{Introduction}
\subsection{Threats by Epidemics}

Sometimes, epidemics management turns out to be unsuccessful.
The plague pandemics in the middle ages, for example, were not handled
effectively \cite{dean2015,nutton2007}. We have learned much in the meantime.
Despite of all medical progress, however, the recent outbreak of H1N1 
\cite{css2010,oxford2000,zhangchen2009,zhouguo2012} and other infectious 
diseases made clear that biological contagions are still a significant 
threat. The recent Ebola epidemics \cite{osy2015,tulenko2014} has shown, 
that even a known pathogen can go out of control easily 
\cite{philipsmarkham2014}. The recent series of sporadic cases of 
infectious illnesses \cite{colizzaetal2005} is the reason why epidemics 
research is still of high actual importance \cite{ks2014}. This threat 
will persist in the future, eventually even increasing due to a variety 
of effects including evolution of relevant contagions, climate and 
ecosystem change, land use, and increasing travel activities \cite{ks2014}. 

\begin{figure}[tbhp]
\begin{center}
\begin{tikzpicture}[scale=0.3, every node/.style={transform shape}]
\node [draw, rounded corners, fill=black!7,minimum width=6cm,minimum height=2cm,line width=0.5] (v17) at (0,4) {\Large Epidemics Management};
\node [draw, rounded corners, fill=black!7,minimum width=6cm,minimum height=2cm,line width=0.5] (v2) at (0,-3) {\Large Population};
\node [draw, rounded corners, fill=black!7,minimum width=6cm,minimum height=2cm,line width=0.5] (v18) at (0,-7) {\Large Epidemics};
\node [draw, rounded corners, fill=black!7,minimum width=6cm,minimum height=2cm,line width=0.5] (v12) at (0,-11) {\Large Medical Staff};
\node [draw, rounded corners, fill=black!7,minimum width=6cm,minimum height=2cm,line width=0.5] (v8) at (20,-3) {\Large Medical Treatment};
\node [draw, rounded corners, fill=black!7,minimum width=6cm,minimum height=2cm,line width=0.5] (v11) at (-11,-11) {\Large Epidemics Data};
\draw [-{Latex[scale=1.0]}, draw=black, line width=0.5] ([yshift={0.0cm}]v2.west) -- (-5.5,-3) -- (-5.5,-10.6) -- ([yshift={0.4cm}]v11.east);
\draw [-{Latex[scale=1.0]}, draw=black, line width=0.5] ([yshift=-{0.4cm}]v2.west) -- (-5,-3.4) -- (-5,-10.6) -- ([yshift={0.4cm}]v12.west);
\draw [-{Latex[scale=1.0]}, draw=black, line width=0.5] ([yshift=-{0.8cm}]v2.west) -- (-4.5,-3.8) -- (-4.5,-7) -- ([yshift={0.0cm}]v18.west);
\draw [-{Latex[scale=1.0]}, draw=black, line width=0.5] ([yshift={0.4cm}]v12.east) -- (5,-10.6) -- (5,-3.4) -- ([yshift=-{0.4cm}]v2.east);
\draw [-{Latex[scale=1.0]}, draw=black, line width=0.5] ([yshift={0.0cm}]v18.east) -- (4.5,-7) -- (4.5,-3.8) -- ([yshift=-{0.8cm}]v2.east);
\draw [-{Latex[scale=1.0]}, draw=black, line width=0.5] (v2) edge (v8);
\draw [-{Latex[scale=1.0]}, draw=black, line width=0.5] (v11.west) -- (-17,-11) -- (-17,3.6) -- ([yshift=-{0.4cm}]v17.west);
\draw [-{Latex[scale=1.0]}, draw=black, line width=0.5] ([yshift=-{0.4cm}]v11.east) -- ([yshift=-{0.4cm}]v12.west);
\draw [-{Latex[scale=1.0]}, draw=black, line width=0.5] ([xshift={2.4cm}]v2.north) -- ([xshift={2.4cm}]v17.south) node [midway, sloped, above=0.3cm, fill=white] {\Large Observations};
\draw [-{Latex[scale=1.0]}, draw=black, line width=0.5] ([xshift=-{2.4cm}]v17.south) -- ([xshift=-{2.4cm}]v2.north) node [midway, sloped, above=0.3cm, fill=white] {\Large Actions};
\draw [-{Latex[scale=1.0]}, draw=black, line width=0.5] ([xshift={0.4cm}]v8.south) -- (20.4,-16) -- (-11,-16) -- (v11.south);
\draw [-{Latex[scale=1.0]}, draw=black, line width=0.5] (v12.south) -- (0,-15) -- (19.6,-15) -- ([xshift=-{0.4cm}]v8.south);
\end{tikzpicture}
\end{center}
\caption{Graphical representation of an idealized epidemics 
management process as used in traditional approaches to epidemics management.
\label{traditional}}
\end{figure}
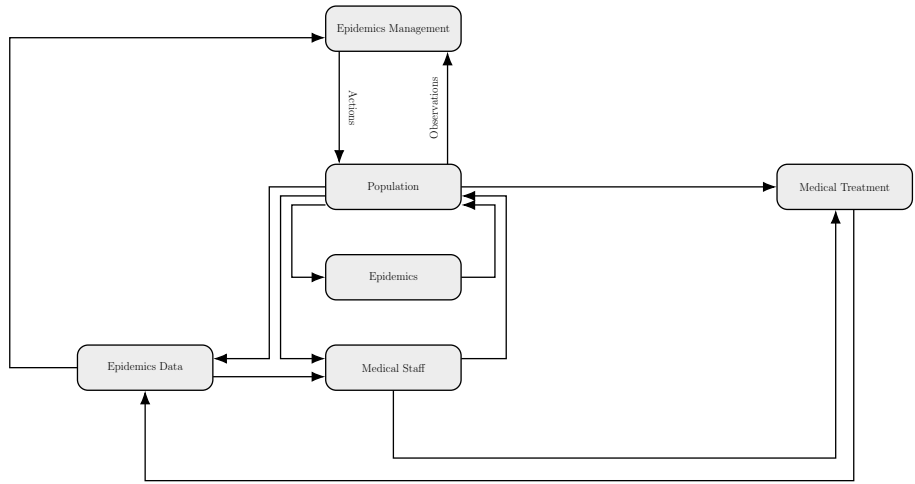

\subsection{Frictions as Threat Multipliers}

The events mentioned in the introduction have shown the vulnerability 
of industrial nations against the threat of known and unknown epidemics.
This has led to the development of various methods for mitigating epidemics 
effects in order to avoid catastrophic developments. Processes and 
organizations have been established with the intention in mind to fight
infectious diseases effectively. Despite of all these efforts, epidemics 
management have been confronted with mishaps and unforeseen problems
with potentially dramatic consequences. For naming but a few examples:
\begin{itemize}
\item The substandard gear used by a Spanish nurse was held responsible for
infecting her with Ebola \cite{kassam2014}
\item Laboratories reported more than 230 safety incidents with 
bioterror viruses and bacteria in 2015 \cite{young2016}. 
\item In 2014, Ebola virus material from a BSL-4 lab leaked out of the lab.
Several people were exposed to the material \cite{muskal2014}.
\item The handling of an U.S. Ebola patient violated many 
regulations \cite{sickles2014}. As an example, it has been forgotten 
to process bedclothes of this patient as highly infectious material 
\cite{dg2014}.
\item An undertaker in Germany infected himself with Lassa fever after
being in contact with the corpse of a Lassa patient who died earlier
\cite{dw2016}.
\item Inefficient communication \cite{hitchcocketal2007,who2015}
leads to delays in distributing decisions and important informations.
\item Especially the deciders themselves, namely the politicians, provided
a whole catalogue of lapses and errors \cite{politics2015}.
\item 
Even a known pathogen may show strongly varying 
properties. For example, Marburg virus \cite{smithetal1967} shows 
variations in lethality from 25\% to 90\% of the infected people
\cite{ndayimirijekindhauser2005,smithetal1967}. 
\item Pathogens are undergoing evolution, leading to unknown and
new, unexpected properties, as for example the 1918 influenza epidemic 
\cite{oxford2000}.
\end{itemize}

According to the examples given above, the management of outbreaks 
of infectious diseases quite often do not work as expected from the 
traditional theories of epidemics and epidemics management.
Their idealistic assumptions are not necessarily valid in practice.
Thus, it seems plausible to include the different kinds of frictions
in the considerations. Frictions will typically influence 
the dynamics of an epidemics outbreak in a more or less significant 
way. Despite of these effects, frictions do not seem to be discussed 
in the necessary depth up to now. The present paper intends to make 
a contribution for closing this gap. Methods for relaxing simplifying 
and idealistic assumptions in epidemics modeling are discussed.

The necessary extensions for a corresponding model will give almost
inevitably a 'complex' model, whereby here the notion of complexity 
has to be understood in the informal sense. Accordingly,
it can be expected that due to the pure size of the 
model, the number of influencing factors, and the number of involved
disciplines the assurance of a high level of objectivity becomes a 
central aspect. This is done from the system 
theory and model analysis point of view. Eventual consequences for the 
medical point of view are not elaborated here.

\subsection{Relevant Papers}

Several papers are considering the interactions between epidemics and 
other areas of science potentially influencing the epidemics like economics
\cite{dutta2008}, human behavior \cite{afh2016,lietal2015,poletti2010},
psychology \cite{epstein2008,papst2015,towersetal2015,wangetal2011},
sociology and knowledge \cite{fast2014,kissetal2010,thomas2012,zelner2011},
and others \cite{righetto2013}. A slightly more complex situation is
considered in \cite{zhangchen2009}. The paper \cite{colizzaetal2007} gives
another view on the complexity applied to epidemics modeling. An article 
supporting a holistic view is \cite{rocaetal2015}. A really 
multidisciplinary perspective on the topic of epidemics or epidemics 
management seems to be still missing, however. 

The statement made above holds for frictions as well. A representation
of a broad range of frictions in a multidisciplinary context appears 
to be a gap in the literature up to now. In the contrary, the paper
\cite{dss2014} shows the necessity of taking such frictions into 
consideration. A paper taking frictions into consideration at least at 
a rudimentary level is \cite{venturiono2016}. Delays as special kind 
of frictions are discussed in \cite{caramagna2011}. Uncertainties were
taken into consideration based on fuzzy logic in \cite{massad2003},
but not by a probability-based representation in a simulation model 
as proposed here.

\subsection{Structure of the Paper}

In section~\ref{complexity}, we analyze the consequences of an inclusion
of frictions, uncertainties and other aspects of nonideality for the
system under consideration. The analysis is done both from the structure 
and dynamics perspective. Based on these results, we derive in the
followoing section~\ref{xyzmodel} stochastic, genericity, and dynamics 
as basic requirements for a model suitably representing such a system. 
As it turns out, an extension of the compartmental modeling approach can 
be considered as appropriate modeling paradigm. Using this paradigm, 
section~\ref{xyzcontrol} discusses the computational complexity of the
epidemics management task. One can safely state, that the computational 
complexity of task for a fully fledged model will usually be not
tractable in practice. Measures of complexity reduction are necessary
for making the problem feasible. The paper closes with a short outlook 
in section~\ref{outlook} summarizing the results and showing options for 
a future development of the topic. 

\section{The System of Epidemics Management}\label{complexity}

For discussing the epidemics management problems we will consider
epidemics management as a system. The exploration of this system provides 
important informations about its properties. They are used in the next 
situation for the decision about an appropriate modeling paradigm.

\subsection{Representation of Influencing Areas}\label{structcomp}

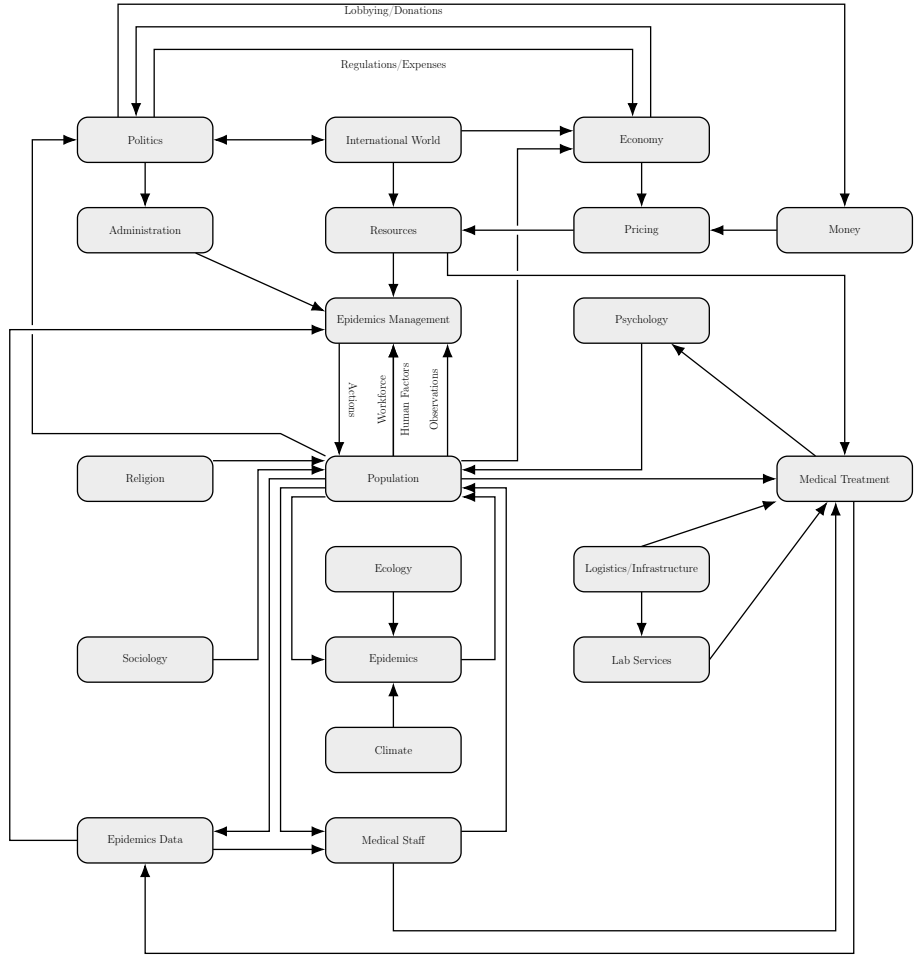
\begin{figure}[htb]
\begin{center}
\begin{tikzpicture}[scale=0.3, every node/.style={transform shape}]
\node [draw, rounded corners, fill=black!7,minimum width=6cm,minimum height=2cm,line width=0.5] (v15) at (0,12) {\Large International World};
\node [draw, rounded corners, fill=black!7,minimum width=6cm,minimum height=2cm,line width=0.5] (v16) at (0,8) {\Large Resources};
\node [draw, rounded corners, fill=black!7,minimum width=6cm,minimum height=2cm,line width=0.5] (v17) at (0,4) {\Large Epidemics Management};
\node [draw, rounded corners, fill=black!7,minimum width=6cm,minimum height=2cm,line width=0.5] (v2) at (0,-3) {\Large Population};
\node [draw, rounded corners, fill=black!7,minimum width=6cm,minimum height=2cm,line width=0.5] (v18a) at (0,-7) {\Large Ecology};
\node [draw, rounded corners, fill=black!7,minimum width=6cm,minimum height=2cm,line width=0.5] (v18) at (0,-11) {\Large Epidemics};
\node [draw, rounded corners, fill=black!7,minimum width=6cm,minimum height=2cm,line width=0.5] (v18b) at (0,-15) {\Large Climate};
\node [draw, rounded corners, fill=black!7,minimum width=6cm,minimum height=2cm,line width=0.5] (v12) at (0,-19) {\Large Medical Staff};
\node [draw, rounded corners, fill=black!7,minimum width=6cm,minimum height=2cm,line width=0.5] (v7) at (11,12) {\Large Economy};
\node [draw, rounded corners, fill=black!7,minimum width=6cm,minimum height=2cm,line width=0.5] (v6) at (11,8) {\Large Pricing};
\node [draw, rounded corners, fill=black!7,minimum width=6cm,minimum height=2cm,line width=0.5] (v19) at (11,4) {\Large Psychology};
\node [draw, rounded corners, fill=black!7,minimum width=6cm,minimum height=2cm,line width=0.5] (v5) at (20,8) {\Large Money};
\node [draw, rounded corners, fill=black!7,minimum width=6cm,minimum height=2cm,line width=0.5] (v8) at (20,-3) {\Large Medical Treatment};
\node [draw, rounded corners, fill=black!7,minimum width=6cm,minimum height=2cm,line width=0.5] (v9) at (11,-7) {\Large Logistics/Infrastructure};
\node [draw, rounded corners, fill=black!7,minimum width=6cm,minimum height=2cm,line width=0.5] (v10) at (11,-11) {\Large Lab Services};
\node [draw, rounded corners, fill=black!7,minimum width=6cm,minimum height=2cm,line width=0.5] (v13) at (-11,12) {\Large Politics};
\node [draw, rounded corners, fill=black!7,minimum width=6cm,minimum height=2cm,line width=0.5] (v14) at (-11,8) {\Large Administration};
\node [draw, rounded corners, fill=black!7,minimum width=6cm,minimum height=2cm,line width=0.5] (v1) at (-11,-3) {\Large Religion};
\node [draw, rounded corners, fill=black!7,minimum width=6cm,minimum height=2cm,line width=0.5] (v3) at (-11,-11) {\Large Sociology};
\node [draw, rounded corners, fill=black!7,minimum width=6cm,minimum height=2cm,line width=0.5] (v11) at (-11,-19) {\Large Epidemics Data};
\draw [draw=black, line width=0.5] ([yshift={0.8cm}]v2.east) -- (5.5,-2.2) -- (5.5,5.8);
\draw [draw=black, line width=0.5] (5.5,6.2) -- (5.5,7.8);
\draw [-{Latex[scale=1.0]}, draw=black, line width=0.5] (5.5,8.2) -- (5.5,11.6) -- ([yshift=-{0.4cm}]v7.west);
\draw [draw=black, line width=0.5] (v2.north west) -- (-5,-1) -- (-16,-1) -- (-16,3.4);
\draw [-{Latex[scale=1.0]}, draw=black, line width=0.5] (-16,3.8) -- (-16,12) -- (v13.west);
\draw [-{Latex[scale=1.0]}, draw=black, line width=0.5] (v6.west) -- (v16.east);
\draw [-{Latex[scale=1.0]}, draw=black, line width=0.5] ([yshift={0.8cm}]v1.east) -- ([yshift={0.8cm}]v2.west);
\draw [-{Latex[scale=1.0]}, draw=black, line width=0.5] (v3.east) -- (-6,-11) -- (-6,-2.6) -- ([yshift={0.4cm}]v2.west);
\draw [-{Latex[scale=1.0]}, draw=black, line width=0.5] ([yshift={0.0cm}]v2.west) -- (-5.5,-3) -- (-5.5,-18.6) -- ([yshift={0.4cm}]v11.east);
\draw [-{Latex[scale=1.0]}, draw=black, line width=0.5] ([yshift=-{0.4cm}]v2.west) -- (-5,-3.4) -- (-5,-18.6) -- ([yshift={0.4cm}]v12.west);
\draw [-{Latex[scale=1.0]}, draw=black, line width=0.5] ([yshift=-{0.8cm}]v2.west) -- (-4.5,-3.8) -- (-4.5,-11) -- ([yshift={0.0cm}]v18.west);
\draw [-{Latex[scale=1.0]}, draw=black, line width=0.5] ([yshift={0.4cm}]v12.east) -- (5,-18.6) -- (5,-3.4) -- ([yshift=-{0.4cm}]v2.east);
\draw [-{Latex[scale=1.0]}, draw=black, line width=0.5] ([yshift={0.0cm}]v18.east) -- (4.5,-11) -- (4.5,-3.8) -- ([yshift=-{0.8cm}]v2.east);
\draw [-{Latex[scale=1.0]}, draw=black, line width=0.5] (v18a) edge (v18);
\draw [-{Latex[scale=1.0]}, draw=black, line width=0.5] (v18b) edge (v18);
\draw [-{Latex[scale=1.0]}, draw=black, line width=0.5] (v5) edge (v6);
\draw [-{Latex[scale=1.0]}, draw=black, line width=0.5] (v7) edge (v6);
\draw [-{Latex[scale=1.0]}, draw=black, line width=0.5] (v9) edge (v10);
\draw [-{Latex[scale=1.0]}, draw=black, line width=0.5] (v9.north) -- (v8);
\draw [-{Latex[scale=1.0]}, draw=black, line width=0.5] (v10.east) -- (v8);
\draw [-{Latex[scale=1.0]}, draw=black, line width=0.5] (v2) edge (v8);
\draw [-{Latex[scale=1.0]}, draw=black, line width=0.5] (v8) edge (v19);
\draw [-{Latex[scale=1.0]}, draw=black, line width=0.5] (v14) -- ([yshift={0.4cm}]v17.west);
\draw [-{Latex[scale=1.0]}, draw=black, line width=0.5] (v11.west) -- (-17,-19) -- (-17,3.6) -- ([yshift=-{0.4cm}]v17.west);
\draw [-{Latex[scale=1.0]}, draw=black, line width=0.5] (v19.south) -- (11,-2.6) -- ([yshift={0.4cm}]v2.east);
\draw [-{Latex[scale=1.0]}, draw=black, line width=0.5] ([yshift=-{0.4cm}]v11.east) -- ([yshift=-{0.4cm}]v12.west);
\draw [-{Latex[scale=1.0]}, draw=black, line width=0.5] (v13) edge (v14);
\draw [-{Latex[scale=1.0]}, draw=black, line width=0.5] (v15) edge (v16);
\draw [-{Latex[scale=1.0]}, draw=black, line width=0.5] (v16) edge (v17);
\draw [-{Latex[scale=1.0]}, draw=black, line width=0.5] ([yshift={0.4cm}]v15.east) -- ([yshift={0.4cm}]v7.west);
\draw [-{Latex[scale=1.0]}, draw=black, line width=0.5] ([xshift={2.4cm}]v16.south) -- (2.4,6) -- (20,6) -- (v8.north);
\draw [-{Latex[scale=1.0]}, draw=black, line width=0.5] ([xshift={2.4cm}]v2.north) -- ([xshift={2.4cm}]v17.south) node [midway, sloped, above=0.3cm, fill=white] {\Large Observations};
\draw [-{Latex[scale=1.0]}, draw=black, line width=0.5] ([xshift=-{2.4cm}]v17.south) -- ([xshift=-{2.4cm}]v2.north) node [midway, sloped, above=0.3cm, fill=white] {\Large Actions};
\draw [-{Latex[scale=1.0]}, draw=black, line width=0.5] (v2) -- (v17) node [midway, sloped, above=0.2cm, fill=white] {\Large Workforce};
\draw [-{Latex[scale=1.0]}, draw=black, line width=0.5] (v2) -- (v17) node [midway, sloped, below=0.2cm, fill=white] {\Large Human Factors};
\draw [{Latex[scale=1.0]}-{Latex[scale=1.0]}, draw=black, line width=0.5] (v13) edge (v15);
\draw [-{Latex[scale=1.0]}, draw=black, line width=0.5] ([xshift={0.4cm}]v8.south) -- (20.4,-24) -- (-11,-24) -- (v11.south);
\draw [-{Latex[scale=1.0]}, draw=black, line width=0.5] (v12.south) -- (0,-23) -- (19.6,-23) -- ([xshift=-{0.4cm}]v8.south);
\draw [-{Latex[scale=1.0]}, draw=black, line width=0.5] ([xshift={0.4cm}]v13.north) -- (-10.6,16) -- (10.6,16) node [midway, below=0.3cm, fill=white] {\Large Regulations/Expenses} -- ([xshift=-{0.4cm}]v7.north);
\draw [-{Latex[scale=1.0]}, draw=black, line width=0.5] ([xshift={0.4cm}]v7.north) -- (11.4,17) -- (-11.4,17) node [midway, above=0.3cm, fill=white] {\Large Lobbying/Donations} -- ([xshift=-{0.4cm}]v13.north);
\draw [-{Latex[scale=1.0]}, draw=black, line width=0.5] ([xshift=-{1.2cm}]v13.north) -- (-12.2,18) -- (20,18) -- (v5.north);
\end{tikzpicture}
\end{center}
\caption{Figure~\ref{traditional} supplemented by important disciplines 
and components with a potential influence on the epidemics management 
process.\label{therelationships}}
\end{figure}

The mathematical modeling of epidemics and actions for their management 
has reached an advanced state \cite{brauer2008,getzlloydsmith2006,
hethcote2000}. The overwhelming majority of the scientific literature 
deals with idealized situations, however. If we want to provide a model which 
conveys the complications of epidemics management resulting from practice,
it will be necessary to represent different kinds of frictions. In many
cases this intention will lead to an ambitious project, because it may 
require the modeling of both the influences causing frictions as well as 
the organizational processes for resolving them. These model extensions
have to be integrated with the epidemics management core model and will
lead almost inevitably to a multidisciplinary model describing
the relationships, interactions, and collaboration between many
epidemics-relevant disciplines \cite{la2013,rghf2014}. Let us take a 
look at the diversity of potentially epidemics-relevant disciplines as
seen in figure~\ref{therelationships}.
\begin{itemize}
\item Ecology may become important as soon as hosts and vectors 
are important for spreading a disease \cite{ebc2016,wr2006}
\item Climate may have a significant influence on the spreading of
an epidemics \cite{desclouxetal2012,hayetal2005}.
\item Sociological aspects like traditional practices and burial rites 
\cite{tulenko2014,undp2014} are influencing the infection rate.
\item Logistics and infrastructure are important e.g. for distributing 
vaccination sets and transporting infected persons (and infectious 
material) to hospitals and quarantine units \cite{eggletonogilvie2010}.
\item Politics, including different handling of different ethnic groups,
opportunism, and corruption \cite{politics2015}, is involved at the 
global level. In effect, politics may also be responsible for wars;
actually, Syrian civil war is taking place causing streams of refugees.
The refugees may be ill, eventually triggering epidemics or help 
spreading them \cite{alawiehetal2014,sk2014}.
Furthermore, governments are responsible for a suitable preparedness 
and a suitable execution of epidemics management actions.
\item The administration defines regulations, gives advice, and provides
resources, which are usually limited.
\item International Relationships in both political and medical respect
may influence the admissible or demanded management actions. 
\item Psychology \cite{flynn2010,rghf2014,tulenko2014,ebola2014} make
humans react in different ways and not necessarily according to the 
intended aims of epidemics management
\cite{funksalathejansen2010,lietal2015,poletti2010}.
\item The economy is typically sensitively influenced by epidemics, 
since infected people do not belong to the workforce anymore \cite{undp2014}.
\begin{itemize}
\item The costs even of common influenza are enormous
\cite{molinarietal2007}. For the United States, the medical resp. overall 
costs amount to \$10.4 resp. \$87.1 billion in 2003.
\item The recent Ebola epidemics \cite{financing2015,undp2014} was causing 
a transition from economic growth to recession in some African 
states \cite{mullan2015}.
\end{itemize}
Besides of that, the limited resource of money is usually restricting the
possible actions against an epidemics.
\item Religion may have a strong influence (e.g. concerning usage of condoms)
on sexually transmitted epidemics \cite{shawelbassel2014}. Sometimes,
vaccinations are rejected due to religious reasons as well
\cite{murakamietal2014}. Another example were the fears of spreading MERS
during hajj and pilgrimage in Saudi Arabia 2014 \cite{lessler2014}.
Insofar, religion is considered as item of its own, though it belongs 
in principle to the discipline of sociology.
\end{itemize}
Taking this diversity of additional influences into account
and giving up the idealizations of classical epidemics management leads
to a realization of a multidisciplinary model, which will in turn
typically give a model with significant structural complexity.

Furthermore, we can state here about the influences
on the epidemics system. Even worldwide organizations exist with 
the responsibility for handling epidemics as for example the WHO, but 
they can make only recommendations. They are not in the position to 
enforce any kind of regulations. 
Low-level deciders like family doctors, on the other hand, may have
a decisive influence on epidemics dynamics especially in the outbreak phase.
A single mishap at the single-person level --- say a person with a highly
infectious disease not recognized by a physician --- may be amplified 
by epidemics spreading across the population to a system-wide problem. 
Summing up, we are considering a multi-scale system here.

\subsection{Representation of Dynamic Processes}\label{rodp}

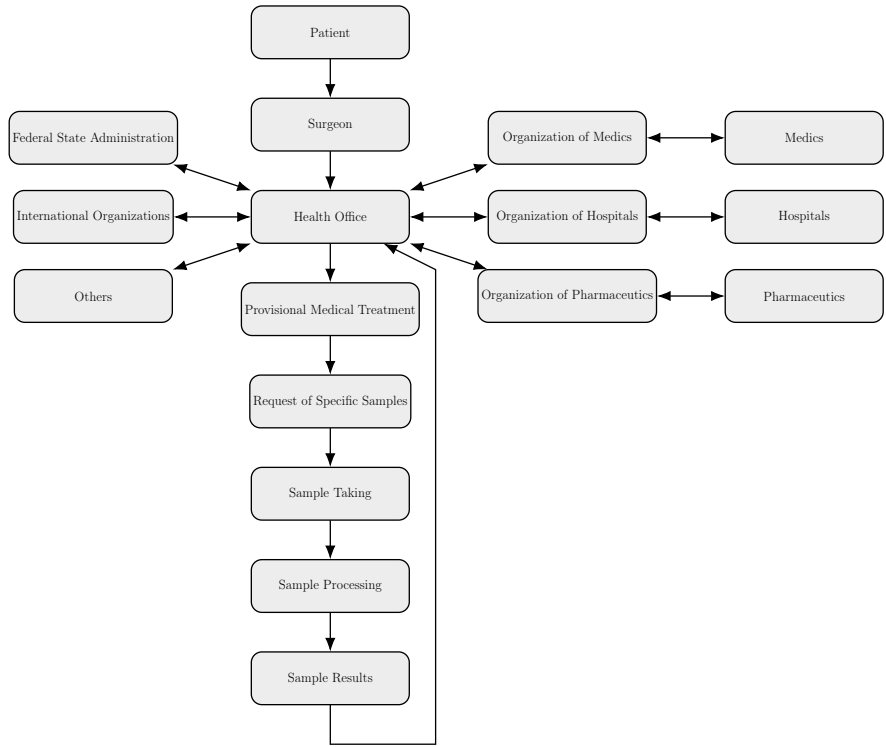
\begin{figure}[tb]
\begin{center}
\begin{tikzpicture}[scale=0.35, every node/.style={transform shape}]
\node [draw, rounded corners, fill=black!7,minimum width=6cm,minimum height=2cm,line width=0.5] (v1) at (0,0) {\Large Patient};
\node [draw, rounded corners, fill=black!7,minimum width=6cm,minimum height=2cm,line width=0.5] (v2) at (0,-3.5) {\Large Surgeon};
\node [draw, rounded corners, fill=black!7,minimum width=6cm,minimum height=2cm,line width=0.5] (v3) at (0,-7) {\Large Health Office};
\node [draw, rounded corners, fill=black!7,minimum width=6cm,minimum height=2cm,line width=0.5] (v4) at (0,-10.5) {\Large Provisional Medical Treatment};
\node [draw, rounded corners, fill=black!7,minimum width=6cm,minimum height=2cm,line width=0.5] (v5) at (0,-14) {\Large Request of Specific Samples};
\node [draw, rounded corners, fill=black!7,minimum width=6cm,minimum height=2cm,line width=0.5] (v6) at (0,-17.5) {\Large Sample Taking};
\node [draw, rounded corners, fill=black!7,minimum width=6cm,minimum height=2cm,line width=0.5] (v7) at (0,-21) {\Large Sample Processing};
\node [draw, rounded corners, fill=black!7,minimum width=6cm,minimum height=2cm,line width=0.5] (v8) at (0,-24.5) {\Large Sample Results};
\node [draw, rounded corners, fill=black!7,minimum width=6cm,minimum height=2cm,line width=0.5] (v9) at (-9,-4) {\Large Federal State Administration};
\node [draw, rounded corners, fill=black!7,minimum width=6cm,minimum height=2cm,line width=0.5] (v10) at (-9,-7) {\Large International Organizations};
\node [draw, rounded corners, fill=black!7,minimum width=6cm,minimum height=2cm,line width=0.5] (v11) at (-9,-10) {\Large Others};
\node [draw, rounded corners, fill=black!7,minimum width=6cm,minimum height=2cm,line width=0.5] (v12) at (9,-4) {\Large Organization of Medics};
\node [draw, rounded corners, fill=black!7,minimum width=6cm,minimum height=2cm,line width=0.5] (v13) at (9,-7) {\Large Organization of Hospitals};
\node [draw, rounded corners, fill=black!7,minimum width=6cm,minimum height=2cm,line width=0.5] (v14) at (9,-10) {\Large Organization of Pharmaceutics};
\node [draw, rounded corners, fill=black!7,minimum width=6cm,minimum height=2cm,line width=0.5] (v15) at (18,-4) {\Large Medics};
\node [draw, rounded corners, fill=black!7,minimum width=6cm,minimum height=2cm,line width=0.5] (v16) at (18,-7) {\Large Hospitals};
\node [draw, rounded corners, fill=black!7,minimum width=6cm,minimum height=2cm,line width=0.5] (v17) at (18,-10) {\Large Pharmaceutics};
\draw [-{Latex[scale=1.0]}, draw=black, line width=0.5] (v1) edge (v2);
\draw [-{Latex[scale=1.0]}, draw=black, line width=0.5] (v2) edge (v3);
\draw [-{Latex[scale=1.0]}, draw=black, line width=0.5] (v3) edge (v4);
\draw [-{Latex[scale=1.0]}, draw=black, line width=0.5] (v4) edge (v5);
\draw [-{Latex[scale=1.0]}, draw=black, line width=0.5] (v5) edge (v6);
\draw [-{Latex[scale=1.0]}, draw=black, line width=0.5] (v6) edge (v7);
\draw [-{Latex[scale=1.0]}, draw=black, line width=0.5] (v7) edge (v8);
\draw [{Latex[scale=1.0]}-{Latex[scale=1.0]}, draw=black, line width=0.5] (v3) edge (v9);
\draw [{Latex[scale=1.0]}-{Latex[scale=1.0]}, draw=black, line width=0.5] (v3) edge (v10);
\draw [{Latex[scale=1.0]}-{Latex[scale=1.0]}, draw=black, line width=0.5] (v3) edge (v11);
\draw [{Latex[scale=1.0]}-{Latex[scale=1.0]}, draw=black, line width=0.5] (v3) edge (v12);
\draw [{Latex[scale=1.0]}-{Latex[scale=1.0]}, draw=black, line width=0.5] (v3) edge (v13);
\draw [{Latex[scale=1.0]}-{Latex[scale=1.0]}, draw=black, line width=0.5] (v3) edge (v14);
\draw [{Latex[scale=1.0]}-{Latex[scale=1.0]}, draw=black, line width=0.5] (v15) edge (v12);
\draw [{Latex[scale=1.0]}-{Latex[scale=1.0]}, draw=black, line width=0.5] (v16) edge (v13);
\draw [{Latex[scale=1.0]}-{Latex[scale=1.0]}, draw=black, line width=0.5] (v17) edge (v14);
\draw [-{Latex[scale=1.0]}, draw=black, line width=0.5] (v8.south) -- (0,-27) -- (4,-27) -- (4,-9) -- (v3);
\end{tikzpicture}
\end{center}
\caption{A seemingly elementary process like the identification of the
disease of an infected patient may have a considerable complexity in 
reality. The picture shows the proposed organization for handling 
suspected Ebola cases in Germany according to \cite{fevd2016}.
The numerous persons involved in the identification process of the disease 
and the transport of and work with maybe highly infectious material have
the potential to produce many additional infections in the case of 
severe mishaps. Furthermore, the information exchange between the 
various institutions may be subject to disturbances. The high potential
for the occurrence of faults and mishaps makes it advisable to include such
processes in a realistic model of epidemics management. \label{process}}
\end{figure}

A system is not only characterized by its structural properties, but also
by its dynamical characteristics. For an adequate representation of the
corresponding processes in a model, the main components involved in such 
processes must be included. As an example of such a process, consider the 
process describing the identification of an illness (see 
figure~\ref{process}) with the special case of Ebola in mind.

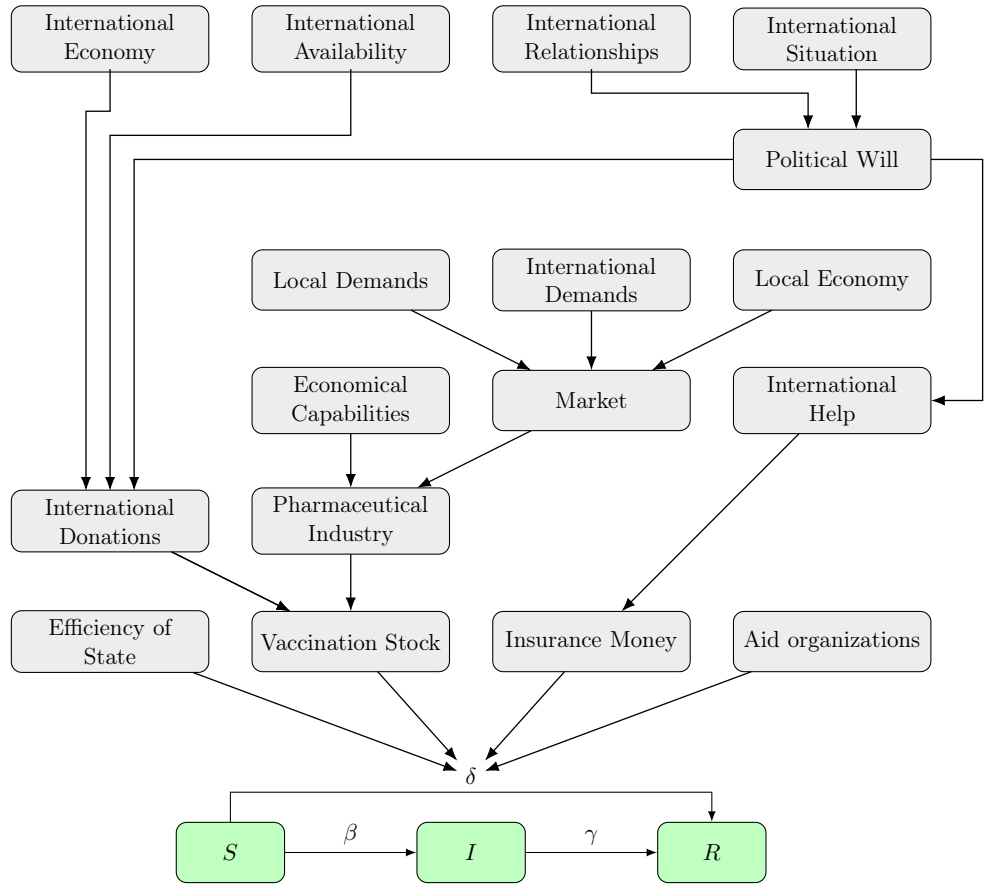
\begin{figure}[tb]
\begin{center}
\begin{tikzpicture}[scale=0.8, every node/.style={transform shape}]
\node [draw, rounded corners, fill=black!7,outer sep=0,inner sep=4,minimum width=3cm, minimum height=1cm] (v2) at (-5,-2.5) {
\begin{minipage}{3.0cm} \begin{center}
Vaccination Stock
\end{center} \end{minipage} };
\node [draw, rounded corners, fill=black!7,outer sep=0,inner sep=4,minimum width=3cm, minimum height=1cm] (v5) at (-1,-2.5) {
\begin{minipage}{3.0cm} \begin{center}
Insurance Money
\end{center} \end{minipage} };
\node [draw, rounded corners, fill=black!7,outer sep=0,inner sep=4,minimum width=3cm, minimum height=1cm] (vxx) at (3,-2.5) {
\begin{minipage}{3.0cm} \begin{center}
Aid organizations
\end{center} \end{minipage} };
\node [draw, rounded corners, fill=black!7,outer sep=0,inner sep=4,minimum width=3cm, minimum height=1cm] (vxy) at (-9,-2.5) {
\begin{minipage}{3.0cm} \begin{center}
Efficiency of State
\end{center} \end{minipage} };
\node [draw, rounded corners, fill=black!7,outer sep=0,inner sep=4,minimum width=3cm, minimum height=1cm] (v1) at (-9,-0.5) {
\begin{minipage}{3.0cm} \begin{center}
International Donations
\end{center} \end{minipage} };
\node [draw, rounded corners, fill=black!7,outer sep=0,inner sep=4,minimum width=3cm, minimum height=1cm] (v3) at (-5,-0.5) {
\begin{minipage}{3.0cm} \begin{center}
Pharmaceutical Industry
\end{center} \end{minipage} };
\node [draw, rounded corners, fill=black!7,outer sep=0,inner sep=4,minimum width=3cm, minimum height=1cm] (v7) at (-5,1.5) {
\begin{minipage}{3.0cm} \begin{center}
Economical Capabilities
\end{center} \end{minipage} };
\node [draw, rounded corners, fill=black!7,outer sep=0,inner sep=4,minimum width=3cm, minimum height=1cm] (v6) at (-1,1.5) {
\begin{minipage}{3.0cm} \begin{center}
Market
\end{center} \end{minipage} };
\node [draw, rounded corners, fill=black!7,outer sep=0,inner sep=4,minimum width=3cm, minimum height=1cm] (v4) at (3,1.5) {
\begin{minipage}{3.0cm} \begin{center}
International Help
\end{center} \end{minipage} };
\node [draw, rounded corners, fill=black!7,outer sep=0,inner sep=4,minimum width=3cm, minimum height=1cm] (v8) at (-5,3.5) {
\begin{minipage}{3.0cm} \begin{center}
Local Demands
\end{center} \end{minipage} };
\node [draw, rounded corners, fill=black!7,outer sep=0,inner sep=4,minimum width=3cm, minimum height=1cm] (v9) at (-1,3.5) {
\begin{minipage}{3.0cm} \begin{center}
International Demands
\end{center} \end{minipage} };
\node [draw, rounded corners, fill=black!7,outer sep=0,inner sep=4,minimum width=3cm, minimum height=1cm] (v10) at (3,3.5) {
\begin{minipage}{3.0cm} \begin{center}
Local Economy
\end{center} \end{minipage} };
\node [draw, rounded corners, fill=black!7,outer sep=0,inner sep=4,minimum width=3cm, minimum height=1cm] (pw) at (3,5.5) {
\begin{minipage}{3.0cm} \begin{center}
Political Will
\end{center} \end{minipage} };
\node [draw, rounded corners, fill=black!7,outer sep=0,inner sep=4,minimum width=3cm, minimum height=1cm] at (3,7.5) {
\begin{minipage}{3.0cm} \begin{center}
International Situation
\end{center} \end{minipage} };
\node [draw, rounded corners, fill=black!7,outer sep=0,inner sep=4,minimum width=3cm, minimum height=1cm] at (-1,7.5) {
\begin{minipage}{3.0cm} \begin{center}
International Relationships
\end{center} \end{minipage} };
\node [draw, rounded corners, fill=black!7,outer sep=0,inner sep=4,minimum width=3cm, minimum height=1cm] at (-5,7.5) {
\begin{minipage}{3.0cm} \begin{center}
International Availability
\end{center} \end{minipage} };
\node [draw, rounded corners, fill=black!7,outer sep=0,inner sep=4,minimum width=3cm, minimum height=1cm] at (-9,7.5) {
\begin{minipage}{3.0cm} \begin{center}
International Economy
\end{center} \end{minipage} };
\draw [default_conn] (v1) edge (v2);
\draw [default_conn] (v1) edge (v2);
\draw [default_conn] (v3) edge (v2);
\draw [default_conn] (v4) edge (v5);
\draw [default_conn] (v6) edge (v3);
\draw [default_conn] (v7) edge (v3);
\draw [default_conn] (v8) edge (v6);
\draw [default_conn] (v9) edge (v6);
\draw [default_conn] (v10) edge (v6);
\draw [default_conn](pw.east) -- (5.5,5.5) -- (5.5,1.5) -- (v4.east);
\draw [default_conn](3.4,7) -- (3.4,6);
\draw [default_conn](-1,7) -- (-1,6.6) -- (2.6,6.6) -- (2.6,6);
\draw [default_conn](pw.west) -- (-8.6,5.5) -- (-8.6,0);
\draw [default_conn](-5,7) -- (-5,5.9) -- (-9,5.9) -- (-9,0);
\draw [default_conn](-9,7) -- (-9,6.3) -- (-9.4,6.3) -- (-9.4,0);
\draw (-7,-6) node[fill=green!25,draw=black, rounded corners, minimum width=1.8cm, minimum height=1cm] (S) {$S$};
\draw (-3,-6) node[fill=green!25,draw=black, rounded corners, minimum width=1.8cm, minimum height=1cm]   (I) {$I$};
\draw (1,-6) node[fill=green!25,draw=black, rounded corners, minimum width=1.8cm, minimum height=1cm] (R) {$R$};
\draw[->] (S) -- (I) node[midway,draw=none,above] {$\beta$};
\draw[->] (I) -- (R) node[midway,draw=none,above] {$\gamma$};
\draw[->] (S.north) -- (-7,-5) -- (1,-5) node[midway,draw=none,above] (da) {$\delta$} -- (R.north);
\draw [default_conn] (v2) edge (da);
\draw [default_conn] (v5) edge (da);
\draw [default_conn] (vxx) edge (da);
\draw [default_conn] (vxy) edge (da);
\end{tikzpicture}
\end{center}
\caption{The vaccination rate $\delta$ is not only determined by the 
capabilities of executing a vaccinations, but also by available vaccination 
sets. They may be at hand in the stocks, provided by donations and bought 
using money coming from the government, international aid, or a bio-insurance.
Taking such influences into consideration is important for assessing the
effectiveness of vaccination as epidemics management action.
\label{influences}}
\end{figure}

We will consider another example from a slightly different perspective.
Sometimes, flow parameters determining system dynamics are influenced
by a large spectra of factors. In first approximation, the infection 
rate $\beta$ is determined by the infectivity of the disease and the
contact behavior between susceptible and infected persons. This situation
in the traditional epidemics management may look different when 
considered from a more detailed perspective (see figure~\ref{influences}).

The two examples given above adumbrate that the dynamics of epidemics 
management is complex and can not be described by a simple formula. 
Indeed, when taking a closer look one will be able to identify 
many feedback loops:
\begin{itemize}
\item Parts of the population are needed for executing epidemics 
management actions. Since people exercising their job become more and
more rare during a serious epidemics, it will be hard to realize the 
intended countermeasures at some point. 
\item Obviously, frictions, countermeasures considered as inefficient, or 
an epidemics seemingly out of control may easily cause strong emotions
like mistrust or fear in the public. This, in turn, can lead to additional
infections due to phenomena like circumventing control points,
rejection of vaccinations, mass flight, or violating quarantine regulations. 
\item An epidemics insurance \cite{financing2015,insurance2016} intends
to mitigate the epidemics effects by providing the resources necessary 
for fighting it. Recognizing the reduction of risks by the insurance, the 
insurance taker may be tempted to reduce own preventive measures.
\end{itemize}
Besides of feedback loops, the epidemics management system contains
trade-offs as well. In such a case, some of the various factors influencing 
epidemics management are compensating each other. Take for example the
question when to start countermeasures. An early start of countermeasures 
will fight the outbreak more effectively. Due to possible disadvantages 
like medical problems caused by vaccinations, or the restrictions of 
personal freedom associated with quarantine regulations, the population 
may partially reject early countermeasures especially in the case of 
initially low infection numbers. In general, the presence of a 
trade-off means that typically not even the tendency of system reaction 
can be predicted. Perturbations may be amplified as well as damped out.

\subsection{Representation of Friction Effects}\label{rofe}

\begin{figure}[tbh]
\begin{center}
\begin{tikzpicture}[scale=0.5, every node/.style={transform shape}]
\node [draw, rounded corners, fill=black!7,minimum width=5cm,minimum height=2cm,line width=0.5] (act) at (-5,9) {\Large Actions $A$};
\node [draw, rounded corners, fill=black!7,minimum width=5cm,minimum height=2cm,line width=0.5] (meas) at (5,9) {\Large Obsdervables $O$};
\node [draw, rounded corners, fill=black!7,minimum width=5cm,minimum height=2cm,line width=0.5] (pop) at (0,3) {\Large Population $P$};
\node [draw, rounded corners, fill=black!7,minimum width=5cm,minimum height=2cm,line width=0.5] (epiman) at (0,15) {
\begin{minipage}{5.0cm}
\begin{center}
\Large Epidemics\\ Management $M$
\end{center}
\end{minipage}};
\draw [-{Latex[scale=1.0]}, draw=black, line width=0.5] ([yshift={0.8cm}]pop.east) -| (meas);
\draw [-{Latex[scale=1.0]}, draw=black, line width=0.5] (meas) |- (epiman.east);
\draw [-{Latex[scale=1.0]}, draw=black, line width=1] (epiman.west) -| (act);
\draw [-{Latex[scale=1.0]}, draw=black, line width=1] (act) |- ([yshift={0.8cm}]pop.west);
%
%
\node [draw, rounded corners, fill=black!7,minimum width=5cm,minimum height=2cm,line width=0.5] (v18) at (0,-0.5) {\Large Epidemics Effects};
\draw [-{Latex[scale=1.0]}, draw=black, line width=0.5] ([yshift=-{0.8cm}]pop.west) -- (-4.5,2.2) -- (-4.5,-0.5) -- ([yshift={0.0cm}]v18.west);
\draw [-{Latex[scale=1.0]}, draw=black, line width=0.5] ([yshift={0.0cm}]v18.east) -- (4.5,-0.5) -- (4.5,2.2) -- ([yshift=-{0.8cm}]pop.east);
\node [dashed,draw,minimum width=6cm,minimum height=2cm,line width=0.5] (inf) at (-12,7) {
\begin{minipage}{7.0cm}
\Large {\bf \color{red} Constraints $C_A$ for $A$}\\[2mm]
\Large Limited Influentiability
\end{minipage}};
\node [dashed,draw,minimum width=6cm,minimum height=2cm,line width=0.5] (obs) at (12,7) {
\begin{minipage}{7.0cm}
\Large {\bf \color{red} Constraints $C_O$ for $O$}\\[2mm]
\Large Limited Observability
\end{minipage}};
\node [dashed,draw,minimum width=6cm,minimum height=2cm,line width=0.5] (noact2) at (-12.0,11) {
\begin{minipage}{7.0cm}
\Large {\bf \color{red} Stochastic Noise $N_A$ for $A$}\\[2mm]
\Large Loss of Resources\\
Adaptation Delay
\end{minipage}};
\node [dashed,draw,minimum width=6cm,minimum height=2cm,line width=0.5] (nomeas2) at (12.0,11) {
\begin{minipage}{7.0cm}
\Large {\bf \color{red} Stochastic Noise $N_O$ for $O$}\\[2mm]
\Large Bad Estimates\\
Uncertainties\\
Missing Informations\\
Measurement Delay
\end{minipage}};
\node [dashed,draw,minimum width=6cm,minimum height=2cm,line width=0.5] (vy) at (12,3) {
\begin{minipage}{7.0cm}
\Large {\bf \color{red} Population $P$ as player}\\[2mm]
\Large Activities, Resistance, \\
\Large Ignorance, Backloops
\end{minipage}
};
\node [dashed,draw,minimum width=6cm,minimum height=2cm,line width=0.5] (vy11) at (-12,3) {
\begin{minipage}{7.0cm}
\Large {\bf \color{red} Other players $\neq M,P$}\\[2mm]
\Large 
Unknowns,
Unexpected Events\\
Changing Situation
\end{minipage}
};
\node [dashed,draw,draw, minimum width=7cm,minimum height=2cm,line width=0.5] (no1) at (-12,22) {
\begin{minipage}{7.0cm}
\Large {\bf \color{red} Strategy $D$}\\[2mm]
\Large 
Approximate Solutions\\
Psychological Influence\\
Beliefs\\
Preoccupation\\
Subjectivity\\
Individual Perceptions\\
Personal Safety \\
Public Opinion
\end{minipage}};
\node [dashed,draw,draw, minimum width=7cm,minimum height=2cm,line width=0.5] (no2) at (0,22) {
\begin{minipage}{7.0cm}
\Large {\bf \color{red} Assessment Function $c$}\\[2mm]
\Large 
Individual Assessments\\
Contradicting Assessments\\
Multiobjective Assessments
\end{minipage}};
\node [dashed,draw,minimum width=6cm,minimum height=2cm,line width=0.5] (noact1) at (12,22) {
\begin{minipage}{7.0cm}
\Large {\bf \color{red} Constraints $C$ of $M$}\\[2mm]
\Large Limited Resources\\
Physiological Limits\\
External Influence
\end{minipage}};
\node [minimum width=6cm,minimum height=2cm,line width=0.5] (noactxxyyzz) at (0,12.7) {
\begin{minipage}{7.0cm}
\begin{center}
\Large \color{red} Decision $D$ for\\
assessment functions $c$\\
under constraints $C$
\end{center}
\end{minipage}};
\draw [dashed, draw=black, line width=0.5] (no1.south) -- (epiman);
\draw [dashed, draw=black, line width=0.5] (no2.south) -- (epiman);
\draw [dashed, draw=black, line width=0.5] (noact1.south) -- (epiman);
\draw [dashed, draw=black, line width=0.5] (inf) -- (act);
\draw [dashed, draw=black, line width=0.5] (noact2) -- (act);
\draw [dashed, draw=black, line width=0.5] (obs) -- (meas);
\draw [dashed, draw=black, line width=0.5] (nomeas2) -- (meas);
\draw [dashed, draw=black, line width=0.5] (vy.west) -- (pop.east);
\draw [dashed, draw=black, line width=0.5] (vy11.east) -- (pop.west);
\end{tikzpicture}
\end{center}
\caption{Influences of Frictions on Epidemics Managmenet Process. 
The many influences and control loops make it difficult to 
design purposeful actions of epidemics management for mitigating
epidemics effects.\label{nonidealcontrol}\label{thefrictions}}
\end{figure}
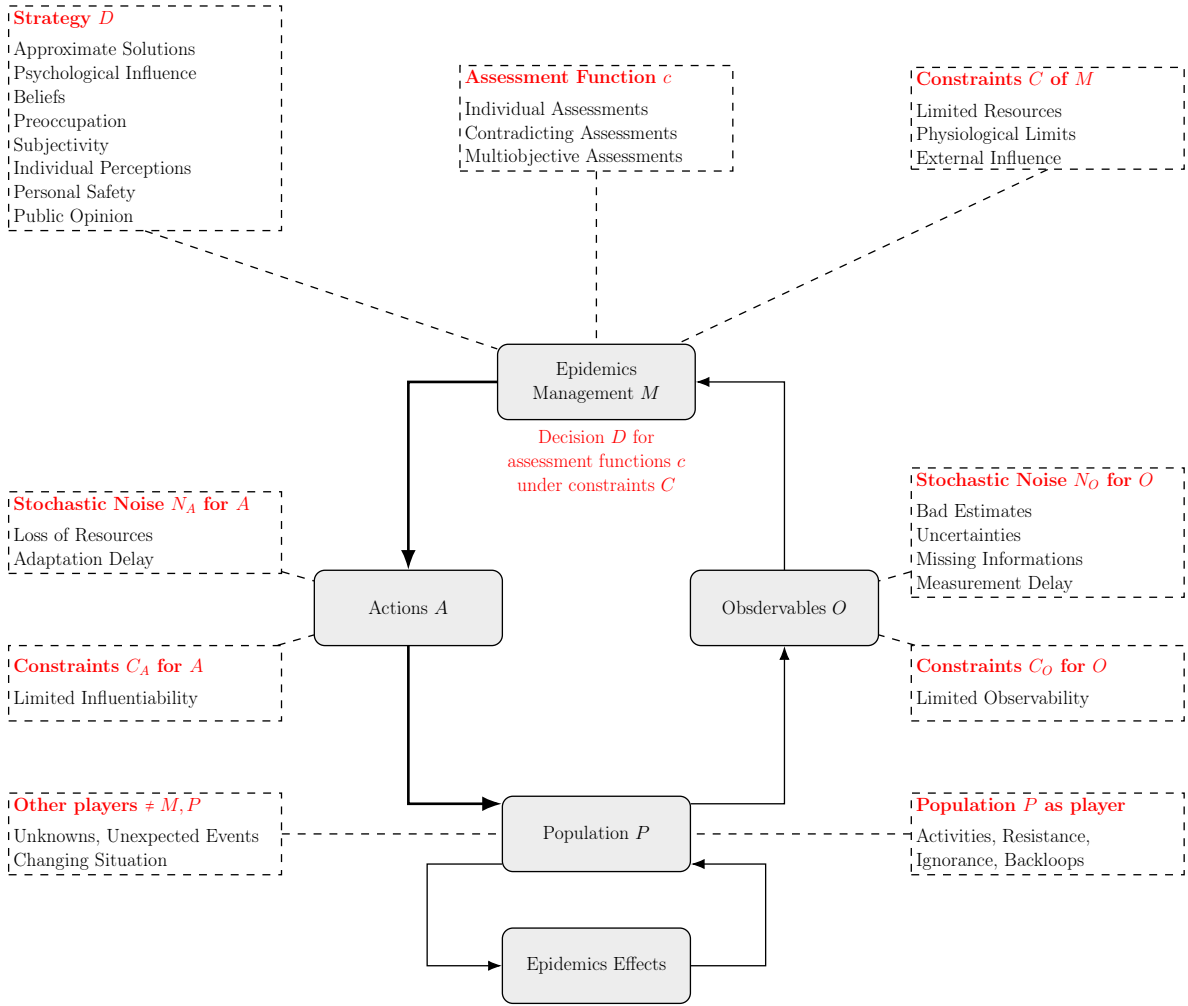

There is a large spectrum of different kinds of frictions
\cite{eggletonogilvie2010,flynn2010,tulenko2014} as shown in 
figure~\ref{thefrictions}. Human factors, delays unknown in advance, 
and unexpected events are occurring. Informations about epidemics are 
becoming available only time after time. Furthermore,
epidemics management is error-prone in principle due to its distributed 
character. Various mistakes can occur during measurement, communication, 
collaboration and so on between the individual control components. Since
these frictions may have a decisive disadvantageous influence on the
outcome of epidemics management \cite{rghf2014,tulenko2014}, their
inclusion seems to be mandatory as soon as a realistic
view on the management process is intended. The following list gives 
a taxonomy of important friction classes together with some explanatory
notes.
\begin{description}[leftmargin=0pt]{}
\item[\rm\em Stochastic Variations, Noise, and Frictions:] In principle, 
the epidemics management process can be disturbed by different kinds of 
noise anywhere and anytime. The noise may be represented by stochastic 
variations leading to erroneous observations results, inadvertently set 
action parameters, inadequate decisions, unforeseen delays, and additional 
waiting times.
\item[\rm\em Observables:] Some observable data may be inaccessible
or missing. The pathogen causing an epidemics may neither always 
be identified without doubt nor may have always known properties. 
Especially in their early stages, misidentification of exotic illnesses 
are quite common. A prominent example is the similarity between the 
symptoms of an early stage of an infection by viral hemorrhagic fevers 
and of an influenza infection \cite{sickles2014}.
\item[\rm\em Multiple Players:] 
All these players have their own interests and aims and trying to
realize their own plans. 
Thus, each authority may make its own decisions.
This may include counteracting the actions of epidemics management.
HIV can serve as an example. Whereas health organizations all over 
the world recommended to use condoms as a protective measure, the pope 
voted against it due to religious reasons. For many people in Africa, 
the pope was the higher authority resulting in dramatic consequences for 
the developing HIV epidemics in Africa \cite{shawelbassel2014}. 
The existence of other players may produce large deviations.
\item[\rm\em Constraints:] 
Selection and application of control actions are prominently influenced 
by constraints. These constraints include the limitation of the available 
resources \cite{isaacs2015,tulenko2014} like medical equipment or mosquito 
nets and of the possible execution rates of management actions like 
vaccinations. Thus, they indirectly influence various parameters of the
epidemics dynamics. 
\item[\rm\em Evaluations:] 
The assessment of the effects of epidemics and epidemics management can be
varied in manifold ways. For example, the set SF36 is a collection of
36 items relevant to physiology and psychology. Another standardized set
of evaluation measures for medicine is PGWB consisting of 22 psychological 
relevant items. The evaluation task is further complicated by different 
personalities and individual perspectives on the analysts, which induces
some kind of subjectivism. Even applying an evaluation measure at different
simulation times may assess different aspects like immediate epidemics 
effects vs. long-term-effects. For naming examples, consider the immediate
impact on the gnp caused by the workforce drop-out vs. the cancellation
of economic investments \cite{financing2015}. Since the different 
evaluation measures are typically incomparable to each other, a 
multiple-criteria evaluation will result.  
\item[\rm\em Actions:]
The dynamics of the epidemics can be influenced by epidemics management 
actions \cite{bs1993,lhscm2007}. Their execution times and parameterizations
are determined by the decisions of epidemics management. 
\end{description}

\section{Models of Epidemics Management}\label{xyzmodel}

\subsection{Requirements for the Modeling Paradigm}

Now, we are going to derive a modeling mechanism, which can represent
the epidemics management system considering our findings about the 
underlying system characteristics. This means that the model must be 
able to handle the existing structural and dynamical complexity. 

At first, we can state immediately that an appropriate model must be
stochastic for representing frictions having a statistic nature as 
stated in section~\ref{rofe}. The simplifications contained in the model 
are another source of stochastic behavior, which are inevitable for
abstracting from the unbounded complications of the real world.
The explicit details of the system omitted in the model are 
transformed to implicit stochastic fluctuations 

Second, the counteracting influences of the epidemics and of the epidemics
management actions together with other complications as indicated in
section~\ref{rodp} --- feedback loops and trade-offs --- gives the
overall system representing epidemics management a complex dynamics. 
A maybe continuous inflow of supplementary informations for the system
components responsible for epidemics management and sporadically occurring 
unexpected events will require an ongoing adaption of the intended 
control actions to the actual situation. Both factors, a complex 
dynamics and ongoing modifications of the course of actions, make an
explicit representation of temporal aspects mandatory. This is done 
by pursuing a simulation-based analysis of the model.

Third, the choice of the model representing the epidemics system may 
quite often be subject to debate. A change of the available level
of informations may require adaptations of the model. Different 
stakeholders may want to analyze the system with different aims in
mind requiring somewhat different models. The deviations between the
models may concern the model parameters, the addition of hosts, 
resources, and vectors, the reproduction of ecological subsystems, or 
the epidemics submodel (e.g. SIR-type vs. SEIR-type). In order to assure 
some kind of comparability of the results provided by different models, 
these models should have a common generic structure.

\subsection{Compartmental Modeling Paradigm}

Summing up the contents of the last section, an adequate modeling 
paradigm provides stochastic simulation models with generic structure.
Compartmental models \cite{braueretal2008,brauer2013,hethcote2000} 
are suitable candidates. As illustration, one can look at the equation 
system describing the basic SIR epidemics model 

\begin{equation}\label{sirequ}
\begin{array}{rclrcl}
\mathrm{d} S / \mathrm{d} t &=& -\beta I\cdot S /(S+I+R) \mathrm{\qquad\qquad\qquad} 
			& S(0) &\ge& 0\\
\mathrm{d} I / \mathrm{d} t &=& \beta I\cdot S /(S+I+R) - \gamma I \mathrm{\qquad\qquad\qquad} 
			& I(0) &\ge& 0\\
\mathrm{d} R / \mathrm{d} t &=& \gamma I
			& R(0) &\ge& 0\\
\end{array}
\end{equation}

wherein the parameters and variables have the following meaning:

\begin{center}
\begin{tabular}{|l|l|}\hline
$S$\qquad\qquad\qquad &
Susceptibles \\ \hline
$I$ &
Infectives \\ \hline
$R$ &
Recovered people with immunity \\ \hline
$\beta$ &
Contact rate \\ \hline
$1 /\gamma$ &
Average infectious period \\ \hline
\end{tabular}
\end{center}

As seen in equation~(\ref{sirequ}), compartmental models have an explicit
time dependence due to their equivalence to a differential resp. difference 
equation. Consequently, the dynamic behavior of the model can be analyzed 
by a simulation of the model. Stochastic variations can be included without 
any problem. Furthermore, compartmental models are generic as well. They 
represent the state of the illness in the population by the 
portion of people being e.g. susceptible, infected, or recovered. This 
concept can be generalized to other attributes like age, sex, job, 
hygienic standards, membership to risk groups etc. A corresponding example 
describing a quite complex situation involving humans, several vectors and 
several hosts is given by the plague \cite{dean2015,nutton2007}. This
flexibility allows to represent system structures as shown in the 
figures~\ref{therelationships}, \ref{process}, and \ref{influences}.

\subsection{Compartmental Models on Networks}

The flexibility of the compartmental modeling paradigm allows to
distinguish different subpopulations provided with individual 
epidemics parameters. This could be realized by using a network $G=(V,E)$
\cite{wittenpoulter2007,yangwandlai2012}, in which the nodes 
$V=\{v_i\}_{i\in I}$ of the network correspond to the different 
subpopulations, whereas the edges $E=\{e_j\}_{j\in J}$ with 
$e_j=(v_1^j,v_2^j)\subseteq V\times V$ correspond to interactions between 
them \cite{apollonietal2013,chowelletal2016,danonetal2011,mr2008}.
This leads to an epidemics dynamics described by the equation system 
\begin{equation}\label{epidemnet}
\begin{array}{rcllrcl}
\mathrm{d} S_i / \mathrm{d} t &=& -\beta_i I_iS_i /(S_i+I_i+R_i) 
	& + \sum_{k\in I, k\neq i} \tau_{ik}^S(t) S_k \mathrm{\qquad\qquad} 
			& S_i(0) &\ge& 0\\
\mathrm{d} I_i / \mathrm{d} t &=& \beta_i I_iS_i /(S_i+I_i+R_i) - \gamma_i I_i
	& + \sum_{k\in I, k\neq i} \tau_{ik}^I(t) I_k 
			& I_i(0) &\ge& 0\\
\mathrm{d} R_i / \mathrm{d} t &=& \gamma_i I_i
	& + \sum_{k\in I, k\neq i} \tau_{ik}^R(t) R_k 
			& R_i(0) &\ge& 0\\
\end{array}
\end{equation}
In the set of equations given above, an index $i$ indicate that the 
corresponding object belongs to the node $v_i\in V$. The flow from the
node $v_i$ to the node $v_k$ for susceptible, infected, and recovered
persons are described by the time-dependent flow parameters 
$1\ge \tau_{ik}^S(t)$, $\tau_{ik}^I(t)$, $\tau_{ik}^R(t) \ge 0$.
Supplementary constraints on the flow parameters assure that
the size of the overall population remains constant over time.
The equation system~(\ref{epidemnet}) is more complex than (\ref{sirequ}),
but can describe specific situations more precisely by e.g. locally 
modified epidemics parameters. This may improve the prediction accuracy.

The compartment approach has the capability of representing single 
persons in principle. Since the behavior of single persons may influence 
the outcome of the epidemics significantly, such a high resolution view
seems to be advantageous according to the observed multi-scale property
of the epidemics system. An agent-based model representing all individuals
of a densely populated nation is intractable from the computational 
complexity point of view, however. 

A decisive advantage of the network approach in this respect is the 
freedom to represent a given situation in different resolutions. A whole 
national state can be modeled as a single SIR model, as a network of 
federal states with edges as neighbourhood relations, as a network of 
roads between towns, villages, airports, hospitals etc. and so on.
This makes it possible to adapt the model to the available computing 
power and the restricted availability of data. It does not make sense to 
use a highly detailed model if the many parameters of such a model can 
not be given specific values by the available data.

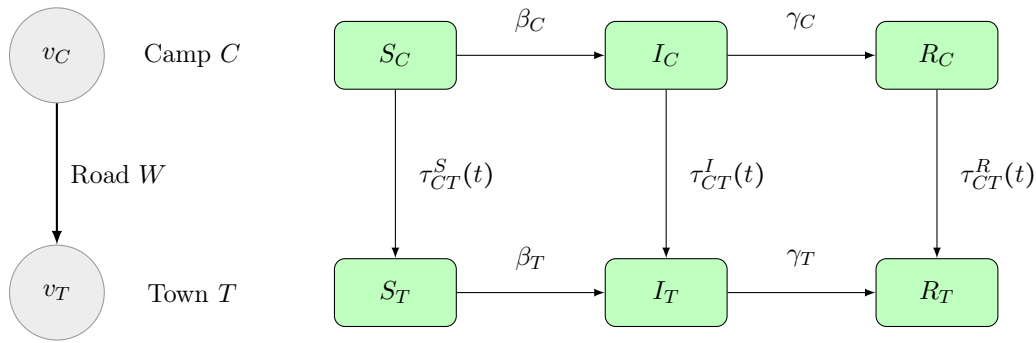
\begin{figure}[tbhp]
\begin{center}
\begin{tikzpicture}[scale=0.9, every node/.style={transform shape,
  draw=black, rounded corners, minimum width=1.8cm, minimum height=1cm},
  line/.style = {->, draw=black, thick},
  box/.style = {circle, draw=black!40, fill=black!7, minimum size=14mm}
								]
    \coordinate (S) at (-5cm, 0cm);
    \coordinate (S1) at (-5cm, -3.5cm);
    \node (Sbox)  [box] at (S)  {$v_C$};
    \node (Sbox1) [box] at (S1) {$v_T$};
    \node[draw=none] at (-3.0,0)  {Camp $C$};
    \node[draw=none] at (-3.0,-3.5) {Town $T$};
    \path[line] (Sbox) -- node [draw=none,right] {Road $W$} (Sbox1);
\draw (0,0) node[fill=green!25] (S) {$S_C$};
\draw (4,0) node[fill=green!25]   (I) {$I_C$};
\draw (8,0) node[fill=green!25] (R) {$R_C$};
\draw[->] (S) -- (I) node[midway,draw=none,above] {$\beta_C$};
\draw[->] (I) -- (R) node[midway,draw=none,above] {$\gamma_C$};
\draw (0,-3.5) node[fill=green!25] (S*) {$S_T$};
\draw (4,-3.5) node[fill=green!25]   (I*) {$I_T$};
\draw (8,-3.5) node[fill=green!25] (R*) {$R_T$};
\draw[->] (S*) -- (I*) node[midway,draw=none,above] {$\beta_T$};
\draw[->] (I*) -- (R*) node[midway,draw=none,above] {$\gamma_T$};
\draw[->] (S) -- (S*) node[midway,draw=none,right] {$\tau_{CT}^S(t)$};
\draw[->] (I) -- (I*) node[midway,draw=none,right] {$\tau_{CT}^I(t)$};
\draw[->] (R) -- (R*) node[midway,draw=none,right] {$\tau_{CT}^R(t)$};
\end{tikzpicture}
\end{center}
\caption{Compartmental model defined on a network representing
a situation consisting of a small camp $C$ and a large town $T$.
Epidemics in $C$ or $T$ is described by a traditional SIR model.
Camp an town are connected by a road $W$, giving a network $G=(V,E)$
consisting of the nodes $V=\{v_C,v_T\}$ and the edge $E=\{W\}$ with
$W=(v_C,v_T)$. People are moving from the camp $C$ to the town $T$ as
described by the flow parameters $\tau_{CT}^S(t)$, $\tau_{CT}^I(t)$, 
$\tau_{CT}^R(t)$.
\label{roadtown}
}
\end{figure}

\subsection{Compartmental Models with Control Components}

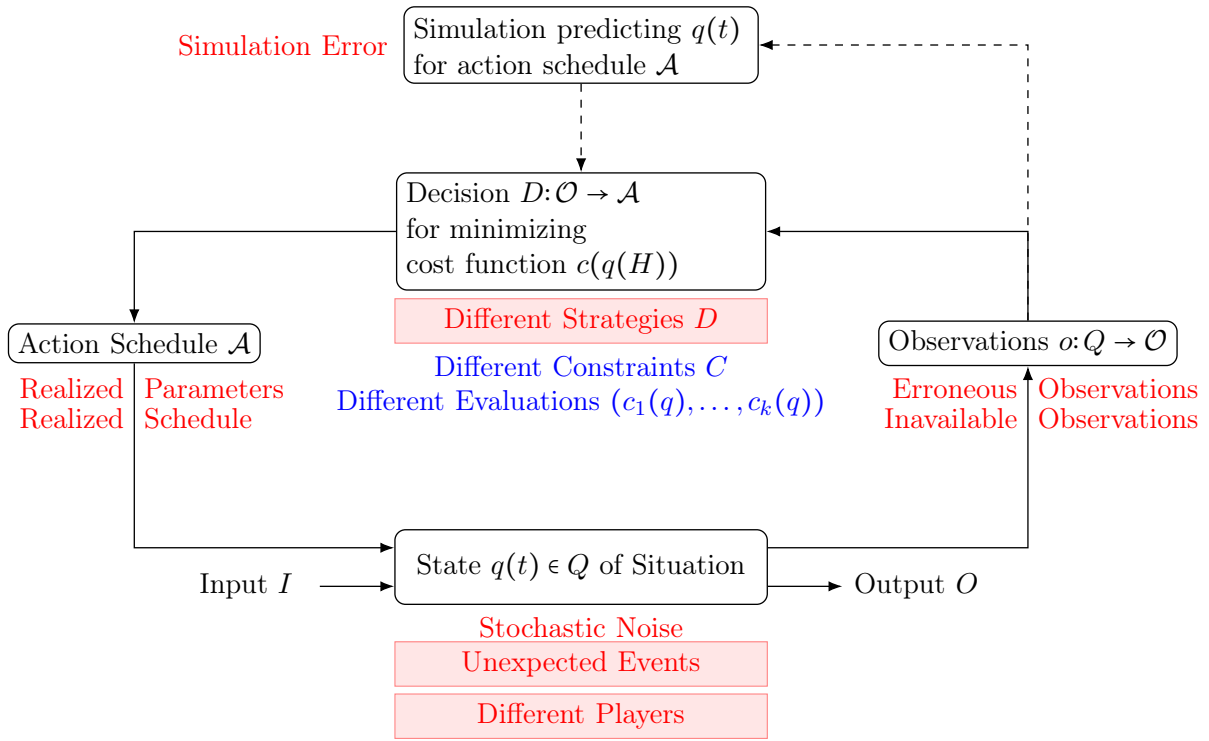
\begin{figure}[tbh]
\begin{center}
\begin{tikzpicture}[scale=0.99, every node/.style={transform shape}]
\node [ draw, minimum width=5cm, minimum height=1cm, rounded corners, line width=0.5] (process) {State $q(t)\in Q$ of Situation};
\node [ draw, yshift=4.5cm, minimum width=4cm, rounded corners, line width=0.5] (control) {
\begin{minipage}{4.7cm}
Decision $D\colon \mathcal{O} \rightarrow \cA$\\
for minimizing\\
cost function $c(q(H))$
\end{minipage}
};
\node [ draw, yshift=7cm, minimum width=4cm, rounded corners, line width=0.5] (sim) {
\begin{minipage}{4.5cm}
Simulation predicting $q(t)$\\
for action schedule $\cA$
\end{minipage}
};
\node [ draw, yshift=3cm, xshift=-6cm, minimum width=2cm, rounded corners, line width=0.5] (actions) {Action Schedule $\cA$};
\node [ draw, yshift=3cm, xshift=6cm, minimum width=2cm, rounded corners, line width=0.5] (indicators) {Observations $o\colon Q\rightarrow
\mathcal{O}$};
\node [ xshift=-4.5cm, yshift=-0.25cm, rounded corners, line width=0.5] (input) {Input $I$};
\node [ xshift=4.5cm, yshift=-0.25cm, rounded corners, line width=0.5] (output) {Output $O$};
\draw[-{Latex[scale=1.0]}, draw=black, line width=0.5] ([yshift=-0.25cm] process.north east) -| (indicators.south);
\draw[-{Latex[scale=1.0]}, draw=black, line width=0.5] (indicators.north) |- (control.east);
\draw[-{Latex[scale=1.0]}, draw=black, line width=0.5,dashed] (indicators.north) |- (sim.east);
\draw[-{Latex[scale=1.0]}, draw=black, line width=0.5,dashed] (sim.south) -- (control.north);
\draw[-{Latex[scale=1.0]}, draw=black, line width=0.5] (control.west) -| (actions.north);
\draw[-{Latex[scale=1.0]}, draw=black, line width=0.5] (actions.south) |- ([yshift=-0.25cm] process.north west);
\draw[{Latex[scale=1.0]}-, draw=black, line width=0.5] ([yshift=+0.25cm] process.south west) -- ++(-1,0);
\draw[-{Latex[scale=1.0]}, draw=black, line width=0.5] ([yshift=+0.25cm] process.south east) -- ++(+1,0);
\node[yshift=2.4cm, xshift=-6cm,left]{\color{red} Realized};
\node[yshift=2.4cm, xshift=-6cm,right]{\color{red} Parameters};
\node[yshift=2.0cm, xshift=-6cm,left]{\color{red} Realized};
\node[yshift=2.0cm, xshift=-6cm,right]{\color{red} Schedule};
\node[yshift=2.4cm, xshift=6cm,left]{\color{red} Erroneous};
\node[yshift=2.4cm, xshift=6cm,right]{\color{red} Observations};
\node[yshift=2.0cm, xshift=6cm,left]{\color{red} Inavailable};
\node[yshift=2.0cm, xshift=6cm,right]{\color{red} Observations};
%
\node[yshift=-0.8cm]{\color{red} Stochastic Noise};
\node[yshift=-1.3cm,draw=red!50, fill=red!10, minimum width=5cm]{\color{red} Unexpected Events};
\node[yshift=-2.0cm,draw=red!50, fill=red!10, minimum width=5cm]{\color{red} Different Players};
\node[yshift=7.0cm,xshift=-4cm]{\color{red} Simulation Error};
\node[yshift=3.3cm,draw=red!50, fill=red!10, minimum width=5cm]{\color{red} Different Strategies $D$};
\node[yshift=2.7cm]{\color{blue} Different Constraints $C$};
\node[yshift=2.2cm]{\color{blue} Different Evaluations $(c_1(q),\ldots,c_k(q))$};
\end{tikzpicture}
\end{center}
\caption{Principle Structure of the epidemics management control loop. 
Red text designates frictions of stochastic nature, whereas red background 
indicates frictions having global effects. Blue text indicates frictions
with deterministic influence.
\label{loopstructure}}
\end{figure}

Compartmental models are a kind of graphical representation of a system 
of differential resp. difference equations. Such a modeling approach 
can describe epidemics dynamics and epidemics effects on the population, 
but it is not considered as appropriate for epidemics management. A 
management process typically involves decisions about a finite number 
of management actions contrary to the continuous world of compartment 
models. Such decisions are typically made based on logical conditions 
over observation data. 
Thus, we have to supplement the modeling paradigm of compartmental models 
\cite{brauer2008} with a paradigm capable of representing logical
reasoning \cite{dimitrovmeyers2010,rahmandadsterman2008}. Similar 
constructions were used in \cite{aslam2015,teoseetal2011}. Formally,
decision makers are represented as control components, which are 
executing observations, make decisions, and schedule epidemics 
management actions as depicted in figure~\ref{loopstructure}. 
More detailed, a control component works in the following way.

The state $q(t)\in Q$ of the epidemics system at time $t$ is narrowed
down by a control compartment using observations $o\colon Q\rightarrow
\mathcal{O}$. The domain $\mathcal{O}$ of the observations $o$ 
may for example concern the levels of 
compartments. Ideally, they allow a precise determination of $q(t)$; 
in reality, usually only a subspace $Q'\subseteq Q$ of $Q$ is observable.
The informations about $q(t)$ provided by the observations $o(q(t))$
are used in a decision strategy $D\colon \mathcal{O} \rightarrow \cA$ 
for creating a plan $\cA$ intended to mitigate the epidemics. The plan 
$\cA$ consists of epidemics management actions like 
vaccinations, information campaigns, calls for social distancing etc. 
to be executed at times $t' \ge t$. These actions may be
parameterized accordingly for adapting to specific infection
rates, vaccination capabilities, available quarantine facilities etc. 

The 'final' outcome of a plan $\cA$ with respect to a time horizon $H$ 
is evaluated using an assessment function $c(q(H))$ applied on the state 
$q(H)\in Q$ reached by the system at the time horizon $H$;
the horizon $H$ determines the time interval $[t,H]$ covered by the 
prognosis of the outcome. Interpreting the assessment function $c$ as cost
function, the control component pursues 'minimal' values of $c(q(H))$
characterized by conditions like a vanishing number of infected persons,
very few persons killed by the epidemics, low economical impact and so on. 
The evaluation measure $c$ defines the aim of the control component,
which selects a plan $\cA$ influencing the dynamics of the epidemics
(hopefully) in such a way, that the outcome $q(H)$ after applying $\cA$
on $q$ is indeed minimizing $c(q(H))$ over the set of all epidemics
management plans.

\subsection{Compartmental Models with Friction Effects}\label{hjkl}

As deduced in section~\ref{rofe}, frictions affecting epidemics 
management --- i.e. the control loop --- should be included in epidemics
models aiming at a more realistic model behavior. In the following, we 
discuss how the different types of frictions identified in 
section~\ref{rofe} can be represented in the model.

\begin{description}[leftmargin=0pt]{}
\item[\rm\em Stochastic Variations:] 
Stochastic noise is a very prominent type of frictions affecting both
input and internal system parameters. It can be represented in the model
as random variable influencing the parameter value. The noise 
characteristics is determined by the stochastic distributions
assigned to the random variable and the function modeling the influence
on the parameter value (say, addition or multiplication). Moreover, 
stochastics may be also used for handling uncertain or even completely 
unknown model parameters. Such uncertainties subsume a variety of 
semantically different effects, which may all be described in the 
framework of stochastics. Typically, one may distinguish uncertainties
of measurements, parameters, and stochastics.
\item[\rm\em Multiple Players:]
As mentioned before, epidemics management is characterized as a distributed 
control consisting of persons with own interests and a variety of 
organizations with not well-defined responsibilities. Consequently, the 
actual epidemics situation may be influenced by actions of multiple 
players. The usage of specific observables and actions, the dependence
on the 'personality' of the decider, and phenomena like limited rationality 
take part in the decision making process $D$. As an example of the effects 
resulting from these complications consider the example given in
figure~\ref{roadtown}. As depicted, we are discussing a situation 
consisting of a camp $C$ and a town $T$ connected by a road $W$. If a 
person in the camp gets infected and an epidemics starts, epidemics 
management conducted by the government may aim at an containment of the 
epidemics in the camp $C$. Accordingly, the government may establish a 
check point at the road connecting $C$ and $T$ for blocking any traffic 
inbetween. The inhabitants of the camp, on the other hand, are in more 
and more danger of getting infected when the epidemics is developing in 
$C$. Being interested in their personal safety, the inhabitants are thus
motivated to flee from the camp $C$ to the town $T$ for avoiding an 
infection. Doing so, they probably transfer the epidemics from $C$ to $T$ 
compromising the strategy of epidemics management. In effect, the decisions 
of parts of the population will thus counteract the decisions of the 
government.
\item[\rm\em Constraints:]
Limitation of rates and resources, rules coming from higher level 
components (e.g. politics), and other restrictions concerning the 
execution of management actions define a set $C$ of constraints on 
the selection of the strategy $D$. The constraints $C$ may change over 
time since e.g. additional resources can be produced and bought.
\item[\rm\em Evaluations:] 
Usually, the evaluation measure $c$ is not a scalar, but a vector $c=
(c_1(x),\ldots,c_k(x))$ of single objectives. Since it is not possible 
in the general to find a state $x$ optimizing all $c_j(x)$ 
simultaneously, one is typically confronted with a set of solution
candidates. It is not obvious how to decide in such a case. Usually 
additional factors come into play at this point like preferences or 
risk attitude of the decider. These factors enable the decider to 
filter options and to compare the remaining choices with each other
from her personal perspective allowing a decision at the end.
Due to the interpretation of $c_j$ as some kind of cost function,
in the following $c_j\ge 0$ is assumed.
\end{description}

\section{Computational Complexity of Epidemics Management}\label{xyzcontrol}

\subsection{Computational Complexity of Simulation-based Analysis}

Epidemics management has a natural interest in the question, which 
management actions are suitable. This can be understood as an optimal 
planning problem concerning the actions to be scheduled for mitigating 
the epidemics effects in the best possible way. In the following, a 
simulation-based approach for solving this problem is given with
$S_H^{\cA}(x)$ as system evolution function for the start state resp. 
parametrization $x$ from the start time to the time $t=H$ under 
inclusion of the scheduled actions $\cA$. Later, we 
will also use the notion $S_H^{f,\cA}(x)$ for the corresponding system 
evolution influenced by frictions $f$. 

As mentioned in section~\ref{hjkl}, we must handle multicriteria 
evaluation measures $c$. Using the abbreviation $y:=S_H^{\cA}(x)$, 
the optimization problem to be considered has the form
\begin{equation}
\bar x := \bigmin\limits_{x=(x_1,\ldots,x_n)\in P} (c_1(y),\ldots,c_k(y))\text{\hphantom{aaaaaaa}with }
P=\bigtimes\limits_{i=1}^n I_i
\label{optiproblem}
\end{equation}
$P$ designates the $n$-dimensional parameter space over which the 
optimization has to be executed. The evaluation measures $c_1,\ldots,c_k$ 
are used as optimization objectives.

Unfortunately, non-smooth aspects make it difficult to predict
the overall behavior of the system and give the optimization problem 
a combinatorial character. Two significant effects are contributing to
the non-smoothness. First, the discrete decisions involved in 
the epidemics management process may induce non-differentiabilities
in the evaluation measure components $c_j$. Even discontinuities 
can not be excluded. Second, the optimization has to be 
executed over maybe less or even unstructured domains like the 
position of check points on the network of streets.
This combinatorial character of the optimization problem is 
disadvantageous for the computational complexity \cite{pr1998}, 
because one can usually not make use of an approximation process 
for finding the optimum anymore. An approximation relies on 
certain smoothness and continuity properties not necessarily given here. 
Without further knowledge about the behavior of the system there is no 
other way than to discretize the problem. One such approach consists
of defining a grid covering the parameter space, making
(\ref{optiproblem}) computational in this way. The grid replaces
the maybe continuous domains $I_i$ by finite point sets $I_i'\subseteq 
I_i$ providing a good 
coverage of $I_i$. The modified optimization problem has the form
\begin{equation}
\bar x' := \bigmin\limits_{x=(x_1,\ldots,x_n)\in P'} (c_1(y),\ldots,c_k(y))\text{\hphantom{aaaaaaa}with }
P'=\bigtimes_{i=1}^n I_i'
\label{gridproblem}
\end{equation}

If the grid introduces $m$ representative points for each of the $n$ 
variables $x_1,\ldots,x_n$, the grid will consist of a total of $m^n$
points of support. If the grid is fine enough, the optimum $\bar x'$
found on this grid can be assumed to be a viable approximation of the
real optimum $\bar x$ meaning $\bar x'\approx \bar x$. Since the 
behavior of the system between the grid points is unknown and any 
attempt to reconstruct intermediate values from the data given 
for the points of support thus
disputable, the validity of the assumed approximation property $\bar x'
\approx \bar x$ can not be guaranteed, however. 
The chances of being valid is increased by choosing a finer grid,
though the higher number of simulation runs required for solving
(\ref{optiproblem}) on a finer grid is a limiting factor for 
a grid refinement.

\subsection{Enabling a Simulation-based Analysis by Complexity Reduction}
\label{compredsec}

The optimization problem~(\ref{optiproblem}) is hard to solve 
because of the typically large number of system parameters. Furthermore,
possible discontinuities e.g. due to decisions about management actions
and the eventual presence of discrete aspects hamper the usage of
approximation methods. Thus, one may be interested in methods for
simplifying the optimization problem as long as they do not compromise
the quality of the solution in an inacceptable way.
Such a method is described in the following based on restricting
the given optimization problem to the essential system parameters.

At first the system parameters $x_i$ are ranked with respect to 
their importance for the optimization problem based on a sensitivity 
analysis. According to \cite{xietal2010}, a sensitivity assessment using 
an isolated analysis of the parameters $x_i$ is attractive
because of its simplicity. If we assume that the domain $I_i$ of the 
$i$-th parameter $x_i$ is an interval $I_i=[a_i,b_i]$, this approach
allows an immediate calculation of the sensitivity coefficients $s_i$ 
according to 
\begin{equation}\label{simplesensitivity}
s_i :={S_H(b_i) - S_H(a_i)\over b_i-a_i} 
\end{equation}
as long as the variations of the outcome $S_H$ of the system evolution 
across $I_i$ is limited. When stronger variations occur, the 
simple measure of sensitivity given in (\ref{simplesensitivity}) has
to be replaced by expressions using additional points of support.
In the case of many system parameters $x_i$, the combined handling of
all $n$ variables using a design of experiment may be more effective.
Though a full factorial design may still need $2^n$
points of support, a fractional design may scale down this number
to $n+1$ points of support corresponding to a computational
complexity of order $O(n)$. This means a reduction of the number of 
simulation runs from the exponential to the linear scale, which makes 
the fractional design comparable to method (\ref{simplesensitivity})
from the viewpoint of computational complexity.
A more detailed discussion of sensitivity analysis applied
to epidemics research can be found in \cite{wuetal2013}.

Using the calculated sensitivity of the parameters $x_i$, we can 
now rank the $x_i$ according to their scale of influence on the
system behavior. The ranking makes it possible to simplify
the optimization problem~(\ref{gridproblem}) by restricting the 
parameter space $P$ to the subspace of parameters influencing the 
outcome sensitively. The other parameters are set to fixed values
$x_j^0\in I_j'$. The choice of $x_j^0$ is uncritical because the
parameter $x_j$ has no sensitive influence on $\bar x'$ according 
to construction. Let us assume w.r.o.g. that the indices 
$1,\ldots,n$ are already ordered according to the ranking and that
only the first $n'\le n$ parameters are included in the simplified
optimization procedure. Consequently, the variables
$x_{n'+1},\ldots,x_{n\vphantom{n'}}$ have to be set to fixed values 
$x_{n'+1}^0 \in I_{n'+1}',\ldots,
x_{n\vphantom{n'}}^0 \in I_{n\vphantom{n'}}'$.
This gives the simplified optimization problem
\begin{equation}
\bar x'' := \bigmin\limits_{x=\left(x_1^{\phantom{0}},\ldots,x_{n'}^{\phantom{0}},x_{n'+1}^0,\ldots,x_{n\vphantom{n'}}^0\right)\in P''} (c_1(y),\ldots,c_k(y))\text{\hphantom{aaaaaaa}with }
P''=\bigtimes\limits_{i=1}^{n'} I_i'
\label{simpleproblem}
\end{equation}
Since the parameters set to fixed values have low sensitivity, their
variation would cause only small disturbances of the result. Thus,
it presumably holds $\bar x''\approx \bar x'$. Together with
$\bar x'\approx \bar x$, this leads to $\bar x''\approx \bar x$.

The number $n'$ of parameters $x_i$ included in the simplified 
optimization problem~(\ref{simpleproblem}) is a trade-off
between the computational tractability of the optimization 
problem~(\ref{simpleproblem}) and the quality of approximation 
of the full problem~(\ref{optiproblem}). Tt is expected that 
the parameters excluded from variation in (\ref{simpleproblem}) 
give only small corrections; thus they may refine but should not 
determine the solution. For validating the approximation 
property one can check whether the parameters $x_{n'+1},\ldots,
x_{n\vphantom{n'}}$ excluded from the optimization process are 
indeed producing only small corrections.
This can be done e.g. by executing corresponding Monte-Carlo runs.

\subsection{Principle Strategy of Friction Analysis}

The strategy described above can solve the basic optimization 
problem of epidemics management. We are continuing the considerations 
for the inclusion of frictions in the optimization process (see 
figure~\ref{nonidealcontrol}). Due to their fundamentally different
character, we distinguish deterministic and nondeterministic 
frictions in the following. Many deterministic frictions can be 
included in the optimization problem without any change of the
formalism. The constraints $C$ and effects of limited rationality,
for example, are canonically represented by a modification of the 
mapping $D$ describing the decision strategy. Not that clear is 
an appropriate strategy for the inclusion of nondeterministic
frictions. We provide methods for handling stochastic disturbances 
representing noise and game-theoretic disturbances caused by 
different, not fully cooperative players.

Concerning stochastic noise, the various influences on the system
dynamics are typically independent from each other and will thus 
usually lead only to local variations. 
Game-theoretic events, in the contrary, may be correlated. The
existence of an 'intelligent' player following a distinct 
strategy may produce a 'systematic' deviation from the intended plan. 
This may provoke a {\em globally} different solution, which is a 
fundamental difference to stochastics. A common property of both kinds
of nondeterministic frictions is that 'punctual' considerations are
inadequate. Instead, considerations based on simulation result
families become of interest, whereby the members of such a family
represent the different potential futures.
Ongoing adaptations of the scheduled epidemics management actions 
for correcting unforeseen or unexpected developments and 
for taking additional and new informations into account are 
contributing to the variation of the outcomes.

The effects of nondeterministic frictions are quantified by risk
and robustness measures assessing the effects of stochastic and 
game-theoretic variations. In effect, these measures can be handled
analogous to the cost estimates provided by the evaluation
measures $c_j$. Consequently, for including them in the
optimization problem, the vector $c=(c_1,\ldots,c_k)$ of evaluation 
criteria is extended by the robustness and risk measures.

A remark is necessary concerning the risk measure. Risk is used 
here for characterizing global variations. Thus, one may be tempted 
to argue that risk is a property of the overall system and not a 
quantity assigned to individual points of support. In fact, however,
the risk is calculated for a specific scenario associated with the
point of support, which is typically determined by the action 
schedule for fighting the epidemics. The measured risk does not 
assess the possible disadvantages associated with the epidemics, 
but the possible disadvantages associated with a specific scenario.
Thus, the risk is measured at each point of support.

The extension of the vector $c$ is essential for taking 
nondeterministic frictions into account. Besides of that, the 
sensitivity analysis procedure may have to be adapted as well. 
Since for each point of support of the discretized optimization
problem~(\ref{gridproblem}) the single deterministic outcome is
replaced by a whole set of outcomes caused by stochastic and/or
game-theoretic variations, one may fit a smooth surface middling 
out the variations. A typical candidate for such a surface is a
quadratic multivariate polynom of the form
$$y= d + \sum_i d_ix_i + \sum_j\sum_{k\le j} d_{jk}x_jx_k.$$
The coefficients of the fitted surface can then be used for
deriving the sensitivity coefficients $s_i$ for the parameters $x_i$.

\subsection{Assessing Local Variations: Robustness Measure}
\label{robusec}

As stated above, nondeterministic frictions are taken into account by 
considering the set of possible futures generated by these frictions. 
At first, we consider the case of stochastic frictions. Due to their
local nature, they can be characterized by the size of deviations
from the deterministic outcome. Here, this quantity is measured by
the robustness of the outcome, since it assesses the stability of the 
selected action schedule $\cA$ with respect to the influence of
frictions. A robust schedule $\cA$ is a plausibility argument for the 
feasibility of $\cA$.

Robustness has been defined in many different ways \cite{alietal2003}. 
For our purposes, we will define robustness as the degree to which 
a system can preserve a given set of system properties with a given 
set $F$ of (stochastic) frictions applied to the system. The term 
'preservation of system properties' means that the cost estimates 
provided by the evaluation measures $c_j$ do not exceed excessively 
the outcome for the friction-free situation. The set $F$ is given by 
the frictions included in the underlying model as e.g. seen in 
figure~\ref{loopstructure}. Though $F$ maybe limited to stochastic 
frictions, it may also include other frictions as well.

Formally~\cite{kitano2007}, the robustness $B_j$ of the system $S$
against a set $F$ of frictions can be described as
\begin{equation}\label{robdef}
B_j = \int_F D(c_j(y_f)) \cdot L(c_j(y_f)) \d f
\end{equation}
with $y_f := S_H^{f,\cA}(x)$ for an action schedule $\cA$ and under 
influence of a friction $f\in F$. The function $L(c_j(y_f))$ gives 
the likelihood density for an evaluation measure value equal to
$c_j(y_f)$. Using the abbreviations $y_f := S_H^{f,\cA}(x)$,
$y:=S_H^{\cA}(x)$, the assessment function $D(c_j(y_f))$ is defined
as follows
$$
D(c_j(y_f)) = \begin{cases}
1 & \text{ for } c_j(y_f) \le c_j(y) \\
0 & \text{ for } c_j(y_f) > c_j(y) + \delta\\
c_j(y)/c_j(y_f) & \text{ for } c_j(y) < c_j(y_f) \le c_j(y) + \delta
\end{cases}$$
Let us take a closer look at the different cases. If $c_j(y_f)$ is less or
equal than $c_j(y)$, then the friction (unexpectedly) improves the outcome 
compared to the friction-free situation. This is considered as perfectly
robust, and thus a value of 1 is assigned. On the other hand, if the
inclusion of the friction $f$ is worsening the outcome by more than
a predefined offset $\delta\ge 0$ compared to the friction-free situation,
a value $D(c_j(y_f))= 0$ indicates the presence of a fundamental problem.
For the remaining third case, which describes a situation between these 
two extremes, the value of $D(c_j(y_f))$ is given by $c_j(y)/c_j(y_f)$.
Since the condition of this case is given by $c_j(y) < c_j(y_f) \le
c_j(y) + \delta$, it holds $c_j(y)/c_j(y_f) \in ]0,1]$. Due to 
$0\le c_j(y) < c_j(y_f)$, the division is well-defined. Summing up,
$D(c_j(y_f))$ returns a relative assessment of the robustness of the
outcome by comparing the evaluation under frictions with the 
corresponding value for the friction-free situation.

The definition of $B_j$ has to be discretized for assuring
computability. Accordingly, the codomain $[0,\infty[$ of the 
components $c_j$ of the
evaluation measure is binned by a partition relation $\sim$
to a countable range of values. Based on this modification,
the robustness given by the integral~(\ref{robdef}) is approximated 
by the sum 
\begin{equation}\label{robcalc}
B = \sum_{Z\in [c_j(y_f)]_\sim} D(Z) L(Z) 
\end{equation}
Robustness is measuring specific effects of stochastics on the 
final outcome. A canonical method for calculating the
robustness~(\ref{robcalc}) is the execution of $N_B$ Monte-Carlo runs.
This means in effect, that for calculating $B_j$ for each
of the $m^n$ points of support a total of $N_B\cdot m^n$ simulation runs 
are necessary. The choice of $N_B$ depends on the required statistical 
significance, the underlying model of variations as for example
stochastic vs. systematic variations, and the structure of the model.
This topic is not discussed here any further.

\subsection{Frictions causing Global Variations: Risk Measure}

Though stochastic variations will usually produce only minor 
deviations from the expected behavior, large deviations may very
well occur with a small probability. Large deviations may also
be caused by other players belonging to the system and pursuing
own interests. The already known robustness measure $B_j$ provides the
fraction of outcomes, which are deviating significantly from the
friction-less outcome due to the inclusion of frictions, but it 
does not provide any statement about the system behavior in the 
case of a large deviation. The epidemics management specialist may 
be very well interested in such statements, however, for deciding 
whether the system behavior in the presence of frictions is either
predominantly good-natured or fatal. For closing this gap, the
determination of a risk measure is recommended here, which covers the 
whole space of possible outcomes.

Formally, the risk \cite{sw2009} is defined as expectation value 
of a loss function over a set of possible hazards for the system.
In our case, the loss function is given by an evaluation measure 
$c_j$ assessing the costs associated with the corresponding outcome 
of the simulation run. The hazards to be taken into account are the
frictions $f\in F$. Thus, the risk is calculated based on the likelihood
$L(y_f)$, that a friction $f$ occurring in a situation with
an intended action schedule $\cA$ produces the costs 
$c_j(y_f)$. This leads to the expression
\begin{equation}\label{riskdef}
R_j = \int_F c_j(y_f) L(c_j(y_f)) \d f
\end{equation}
As in the case of robustness, we have to assure computability 
by discretization. This gives
\begin{equation}\label{riskcomp}
R_j = \sum_{Z\in [c_j(y_f)]_\sim} Z L(Z) 
\end{equation}
For computing the risk, two quantities have to be provided ---
likelihood and costs. Concerning the likelihood, one has to ask, 
which outcomes --- distinguished by values of the evaluation
measure $[c_j(y_f)]_\sim$ --- are produced with which frequency. A
simple approach for answering this question is to execute several 
Monte-Carlo runs. Monte Carlo simulations do not always suffice,
however. So called LPHC-events (Low-Probability High-Consequence) 
may be missed by Monte Carlo runs due to their low probability, but
may be a significant contribution to the overall risk due to their
high criticality. For the handling of LPHC events, special methods 
have been developed \cite{tsunemietal2013}; we will not discuss
this topic any further, though, and assume for reasons of simplicity, 
that the risk is determined by executing a certain number of 
simulation runs. Thus, instead of a single simulation run for
calculating the deterministic evaluation measures $c_j(q)$ a number 
$N_R$ of such runs is necessary at each point of support analogous 
to the robustness measure $B_j$. Summing up, this leads to a
total number of $N_R\cdot m^n$ simulation runs for determining 
$R_j$.

\section{Outlook: Discussion and Advanced Problems}\label{outlook}


For providing a realistic model of epidemics management, many 
complications have to be taken into account. They can be represented
in a hybrid model combining system dynamic aspects and the 
decision-making process of epidemics management. The usage of such a
model for the main task of epidemics management --- optimizing the 
epidemics countermeasures --- is impeded by its complexity. Notably
the stochastics of frictions and their effects contribute to the
complexity, because their inclusion in the assessment is realized 
by extending the corresponding criteria vector by robustness and risk 
measures. This extension is acceptable because complexity reduction 
methods as presented in section~\ref{compredsec} can be applied in
case of need. On the other hand, robustness and risk provide
important additional informations about the stochastics.

Robustness can be understood as the ability of an epidemics management 
policy to tolerate perturbations caused by unforeseen frictions and to 
maintain its effectiveness. The more robust an epidemics policy, the less
probable are necessary adaptations of the policy to changed conditions.
Consequently, this property allows the epidemics manager to assess,
to what extent the many uncertainties of reality are tolerated.
Typically, solutions of an epidemics management problem with large
robustness are preferred.

The risk measure takes the criticalities of disadvantageous outcomes 
(e.g. because of delayed countermeasures due to frictions) weighted by 
its likelihood into account. Since at least in some cases preemptive 
countermeasure may avoid or mitigate such disadvantageous developments, 
risk may be an important tool of epidemics management. Typically,
epidemics management policies with a low risk are preferred.

Based on robustness and risk measures, the paper shows that the
epidemics management problem with inclusion of frictions is tractable
in principle. As was indicated in the introduction, it exists a
large number of application domains potentially affecting epidemics 
management via frictions. This poses some danger to be confronted with
the problems of a world model \cite{pm2012} as the one discussed by the
Club of Rome \cite{meadowsetal1972}.

Systems like Forrester's world 
model can not include all relevant aspects in appropriate detail.
For epidemics management, this means that we will not be able to
make valid predictions under all circumstances. Taking the
responsibility of humans for decision-making as an example, the
complexity of human nature is a principal obstacle for 
predictability. Forecasts can trigger a change in behaviour, humans 
can learn and try alternatives in similar situations, some decisions
do not follow from facts but are characterized by instability and 
randomness. Consequently, the accuracy of predictions is not only 
determined by the quality of the underlying model but by luck as well.
An accurate prediction depends on the fortune to guess the decisions 
made by politicians and other stakeholders and the events occurring in 
the system. Though the higher complexity of the epidemics management 
model considered in this paper can not guarantee correct predictions, 
it increases the chances of a good approximation. It is a tool for 
calculating what-if scenarios, for providing an impression of the 
possible futures, and for showing up eventually dangerous developments
in the system behavior.

\section*{Acknowledgements}
I would like to thank S. Tanasic for many valuable and constructive 
remarks. Additionally, assistance provided by S. Gast and S. Öttl
was appreciated.

\end{document}